\documentclass[aps,onecolumn,preprint,superscriptaddress,nofootinbib,floats]{revtex4}
\usepackage{amsmath,amssymb,color,mathrsfs, graphicx,verbatim,epsfig, bbm, wasysym, axodraw}
\usepackage[hyperfootnotes=false]{hyperref}
\usepackage{slashed}
\allowdisplaybreaks

\def\lsim{\mathrel{\rlap{\lower4pt\hbox{\hskip1pt$\sim$}}
    \raise1pt\hbox{$<$}}}
\def\gsim{\mathrel{\rlap{\lower4pt\hbox{\hskip1pt$\sim$}}
    \raise1pt\hbox{$>$}}} 
\newcommand{\vev}[1]{ \left\langle {#1} \right\rangle }

\newcommand{\be}{\begin{eqnarray}}
\newcommand{\ee}{\end{eqnarray}}

\def\addresses#1#2{\hbox to \hsize{\@tablebox{#1}\hfil\@tablebox{#2}}}
\def\@tablebox#1{\vtop{\hsize=5in \begin{flushleft} #1 \end{flushleft}}}

\def\beq{\begin{equation}}
\def\eeq{\end{equation}}
\def\bit{\begin{itemize}}
\def\eit{\end{itemize}}
\def\beqa{\begin{eqnarray}}
\def\eeqa{\end{eqnarray}}

\def\met{$\displaystyle{\not}E_T$}
\def\vecmet{$\vec{\displaystyle{\not}E}_T$}

\def\PYTHIA{{\tt PYTHIA}}

\def\MadGraph{{\tt MadGraph}}
\def\MadGraph5{{\tt MadGraph5}}

\def\FastJet{{\tt FastJet}}

\def\AFB{$A_{FB}$}

\def\Mtt{M_{t\bar t}}

\def\GA{\Gamma_A}
\def\Ytt{Y_{t\bar t}}

\newcommand{\Pperp}{$P_\perp$}

\newcommand{\Aperpl}{$A_\perp^l$}

\begin{document}

\baselineskip 0.6cm

\begin{titlepage}

\thispagestyle{empty}

\begin{flushright}
\end{flushright}

\begin{center}

\vskip 2cm

{\Large \bf Transverse Top Quark Polarization and the \\ \vspace{1mm} $\mathbf{t\bar t}$ Forward-Backward Asymmetry}

\vskip 1.0cm
{\large  Matthew Baumgart$^1$ and Brock Tweedie$^2$}
\vskip 0.4cm
{\it $^1$ Department of Physics, Carnegie Mellon University, Pittsburgh, PA 15213} \\
{\it $^2$ Physics Department, Boston University, Boston, MA 02215} \\
\vskip 1.2cm

\end{center}

\noindent   The forward-backward asymmetry in top pair production at the Tevatron has long been in tension with the Standard Model prediction.  One of the only viable new physics scenarios capable of explaining this anomaly is an $s$-channel axigluon-like resonance, with the quantum numbers of the gluon but with significant axial couplings to quarks.  While such a resonance can lead to a clear bump or excess in the $t\bar t$ or dijet mass spectra, it may also simply be too broad to cleanly observe.  Here, we point out that broad $t\bar t$ resonances generally lead to net top and antitop polarizations transverse to the production plane.  This polarization is consistent with all discrete spacetime symmetries, and, analogous to the forward-backward asymmetry itself, is absent in QCD at leading order.  Within the parameter space consistent with the asymmetry measurements, the induced polarization can be sizable, and might be observable at the Tevatron or the LHC.

\end{titlepage}

\setcounter{page}{1}

\section{Introduction}
\label{sec:intro}

Because top quarks decay before they have a chance to hadronize, spin effects from their production become imprinted on the angular distributions of their decay products.  This fact can be exploited to craft unique searches for new physics, or to help characterize new physics once it has been discovered~\cite{Bernreuther:1993hq,Beneke:2000hk,Frederix:2007gi,Arai:2007ts,Degrande:2010kt,Cao:2010nw,Baumgart:2011wk,Barger:2011pu,Krohn:2011tw,Bai:2011uk,Falkowski:2011zr,Han:2012fw,Fajfer:2012si,Yang:2012ib,Gabrielli:2012pk,Falkowski:2012cu,Baumgart:2012ay}.  As has often been observed, the helicity of the individual top quarks in $t\bar t$ production from unpolarized hadron collisions is expected to be nearly zero due to the parity-conserving nature of QCD~\cite{Bernreuther:2006vg,Moretti:2006nf}.  An observation of net top helicity in $t\bar t$ would therefore serve as a smoking gun of new physics with parity-violating couplings.  Similarly, the tops should be unpolarized when their spins are measured along the beamline, and in general in any basis within the $t\bar t$ production plane.

In recent years, little attention has been paid to the component of the top quark's polarization perpendicular to the production plane, i.e.~its {\it transverse} polarization~\cite{Dharmaratna:1989jr,Dharmaratna2,Kane:1991bg,Bernreuther:1995cx,Godbole:2006tq}.  This component of the top's polarization is completely allowed by all discrete spacetime symmetries, but is nonetheless absent at leading-order in QCD.  This is because transverse polarization is odd under naive time-reversal, which acts like a discrete symmetry for tree-level amplitudes in T-symmetric theories.  Equivalently, the vanishing tree-level polarization can be seen as an absence of relative complex phases between QCD helicity amplitudes.  To generate the transverse polarization in QCD, we must go to loop-level, where it appears up to only $O(1\%)$ strength in certain regions of production phase space~\cite{Bernreuther:1995cx}.  The small predicted value perhaps explains why there have so far been no attempts to measure it.

A somewhat analogous situation occurs for the $t\bar t$ forward-backward asymmetry (\AFB) in $p\bar p$ collisions.  There, the effect is also absent at leading-order, simply due to the fact that the $q\bar q \to t\bar t$ process proceeds through an unpolarized $s$-channel gluon that carries no information about the spatial orientation of the initial state, and that the $gg \to t\bar t$ process has a symmetric initial-state.  A nonzero \AFB\ is nonetheless generated by box and real emission diagrams.  The inclusive value of \AFB\ is predicted to be about 7\% at the Tevatron~\cite{Almeida:2008ug,Antunano:2007da,Bowen:2005ap,Hollik:2011ps,Kuhn:2011ri,Manohar:2012rs}, which is large enough to measure.  Famously, this is not the value measured by CDF and D0, but instead they observe values above 15\%, about 2--3$\sigma$ high relative to the Standard Model~\cite{Aaltonen:2012it,CDF:AFB2011dilep,Abazov:2011rq,Abazov:2012bfa}.  The interest in explaining this anomaly has generated many new physics models, few of which now survive constraints from the LHC.

Given that \AFB\ and transverse polarization are both allowed effects in QCD that are accidentally small, it is natural to ask whether one may be related to the other within a given new physics scenario, so that the \AFB\ anomaly might imply a transverse polarization anomaly.  Within the present landscape of viable models, this is in fact becoming a likely prospect.  One of the simplest surviving ideas for generating \AFB\ is to introduce a new spin-one color-octet particle with axial couplings to quarks, commonly called an axigluon~\cite{Frampton:1987dn,Hall:1985wz,Frampton:2009rk,Ferrario:2009bz,Tavares:2011zg}.  The $s$-channel exchange of this particle in $q\bar q \to t\bar t$ then interferes with the exchange of a normal gluon and induces the asymmetry, much like the $\gamma$--$Z$ interference at LEP.  In order to avoid direct searches for $t\bar t$ and dijet resonances, it has been proposed that the axigluon is simply too broad to be resolved as a resonance bump~\cite{Tavares:2011zg}.  This substantial width could arise from large couplings to tops or from additional decays into an expanded sector of elusive new colored particles~\cite{Tavares:2011zg,Gross:2012bz}.  The width of a resonance is formally a loop effect, and can therefore induce transverse top polarization in the vicinity of the resonance peak through interference with QCD.  For a general spin-one color-octet, both \AFB\ and the transverse polarization are proportional to the product of axial couplings to top quarks and light quarks, further tightening the relationship.

In this paper, we will quantify the correlation between \AFB\ and transverse polarization within this class of models, and estimate the latter's measurement prospects given the size of the former.  For broad axigluons that are just above $t\bar t$ pair production threshold, we find that the Tevatron may already be in a good position to measure a nonzero inclusive transverse polarization, up to nearly 3$\sigma$ statistical significance.  Heavier axigluons, near or above a TeV, are sometimes considered disfavored by limits from the $t\bar t$ mass spectrum at the LHC, though they may in fact be hidden in exactly the same manner.  The couplings required to explain \AFB\ are larger, leading to a more pronounced resonance peak in $q\bar q \to t\bar t$.  But this peak can easily be beyond the reach of the Tevatron, and obscured by non-resonant $gg\to t\bar t$ at the LHC.  Nonetheless, the transverse polarization near the resonance should be measurable at the LHC, possibly with discovery-level significance in the 2012 data set.  Even the contact-interaction limit, where no clear peak appears at either collider, can lead to observable effects.

Our paper is organized as follows.  In the next section, we describe the transverse polarization expected from QCD, and outline how this component of top polarization can be measured.  In section~\ref{sec:axigluon}, we show in detail the connection between \AFB\ and transverse polarization in axigluon models, and introduce a handful of benchmark models.  In section~\ref{sec:measurement}, we estimate the measurement prospects at the Tevatron and LHC including realistic detector effects and reconstructions.  We also include some supplemental analysis of our benchmark axigluons' effects on $t\bar t$ spin correlations.  We conclude in section~\ref{sec:conclusions}.

\section{Transverse Polarization and QCD}
\label{sec:yPol}

By far the dominant production mechanism for top quarks at both the Tevatron and LHC is through QCD.  Because the hadron beams at these colliders are unpolarized, and because QCD respects the discrete spacetime symmetries P and C, the spins of the emerging top quarks are highly constrained.  Looking at just the top, and tracing out the spin of the antitop (or vice-versa), any net polarization within the production plane is forbidden.  Nonetheless, the polarization component transverse to the production plane (\Pperp) is allowed.  The only constraint comes from charge-conjugation invariance, which requires that the net transverse polarizations of the top and antitop are the same.

To obtain transverse polarization, we must be able to define an oriented production plane.  In the parton-level CM frame for $q\bar q \to t\bar t$, this can be defined by crossing the initiating quark's momentum into the charge +2/3 top's momentum.  For $gg\to t\bar t$, there is no unique choice of axis orientation along the beamline, but a polarization that flips sign between ``forward'' and ``backward'' hemispheres, however defined, is still physically meaningful.  These constructions become ill-defined as we approach the limits of either threshold production or forward production, and \Pperp\ smoothly goes to zero there.  For $gg$, \Pperp\ also vanishes at central production.

At the level of spin amplitudes, transverse polarization requires a specific type of interference.  For definiteness, we can construct a common spin basis starting from the partonic CM frame, by taking the top momentum as the $z$-axis (with basis states $|\!\!\uparrow\rangle$ and $|\!\!\downarrow\rangle$), the above transverse axis as the $y$-axis ($\hat y \parallel \hat{p}(q \; {\rm or} \; g) \times \hat{p}(t)$), and the $x$-axis as the remaining orthogonal direction within the production plane ($\hat x \equiv \hat y \times \hat z$).  We then ``measure'' the spins of the top and antitop after actively boosting them to rest without rotation.  In this basis, \Pperp\ only arises from final-state spin wavefunctions such as $|\!\!\uparrow\rangle \otimes |\!\!\uparrow \rangle + i |\!\!\downarrow\rangle \otimes |\!\!\uparrow\rangle = (|\!\!\uparrow\rangle + i |\!\!\downarrow\rangle) \otimes |\!\!\uparrow\rangle$, where the first ket indicates the spin of the top and the second indicates the spin of the antitop.  At tree-level in QCD, the spin amplitudes in this basis are all relatively real, and \Pperp\ is therefore only generated at loop-level.  This can be viewed as a consequence of the naive time-reversal invariance exhibited by T-symmetric theories at tree-level.\footnote{Naive time-reversal invariance in top pair production follows from the full T-symmetry of QCD and the tree-level equality ${\mathcal M}^{\rm tree}(q\bar q\;{\rm or}\;gg \to t\bar t) = {\mathcal M}^{\rm tree}(t\bar t \to q\bar q\;{\rm or}\;gg)^*$ due to the order-by-order unitarity of the $S$-matrix.  Combined, these result in a tree-level symmetry between processes where the momenta and spins of all initial and final particles are inverted but the initial and final states are not swapped.  We can apply this transformation to top production, and rotate by 180-degrees within the production plane to restore the original momentum configuration.  The symmetry then forces the spin amplitudes for production of $S_y(t) = +1/2$ and $S_y(t) = -1/2$ (with all other particles in fixed helicity eigenstates) to be identical up to a phase, and therefore $\vev{S_y(t)} = 0$.  We will use the same logic below to argue that the apparent transverse polarization induced by backgrounds is also practically zero at leading-order.}  We can also see that \Pperp\ must shut off when the tops are produced with large $p_T$, as contributions from opposite-spin (i.e. same-helicity) states such as $|\!\!\downarrow\rangle \otimes |\!\!\uparrow\rangle$ become suppressed in the chiral limit.

Taken together, these observations give us a schematic form for the magnitude of \Pperp\ in QCD~\cite{Kane:1991bg},
\beq
P_\perp \,\sim\, \alpha_s  \frac{\beta \, m_t}{\Mtt} f(\Theta) \, ,
\label{eq:PperpQCD}
\eeq
where $\beta$ is the top's or antitop's velocity in the partonic CM frame, $\Mtt$ is the $t\bar t$ invariant mass, $\Theta$ is the production angle, and $f$ is a function that vanishes at $\Theta = 0,\pi$.  Because $f$ flips sign between $\Theta$ and $\pi-\Theta$ for $gg\to t\bar t$, any measurement of \Pperp\ that integrates over forward and backward production hemispheres will average out this component of the polarization, leaving over only the $q\bar q\to t\bar t$ contribution.

\begin{figure}[tp]
\begin{center}
\epsfxsize=0.48\textwidth\epsfbox{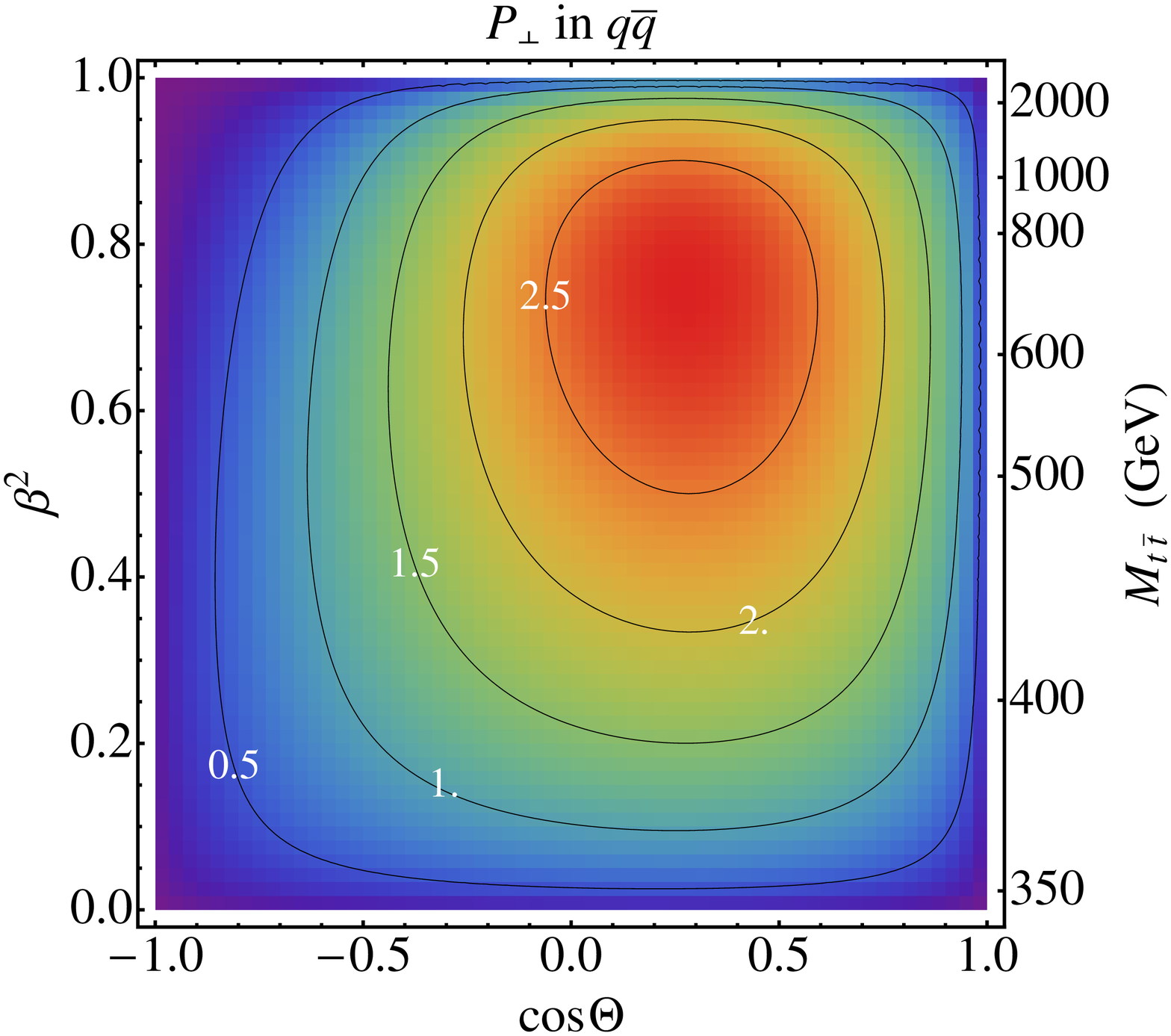} \hspace{0.1in}
\epsfxsize=0.48\textwidth\epsfbox{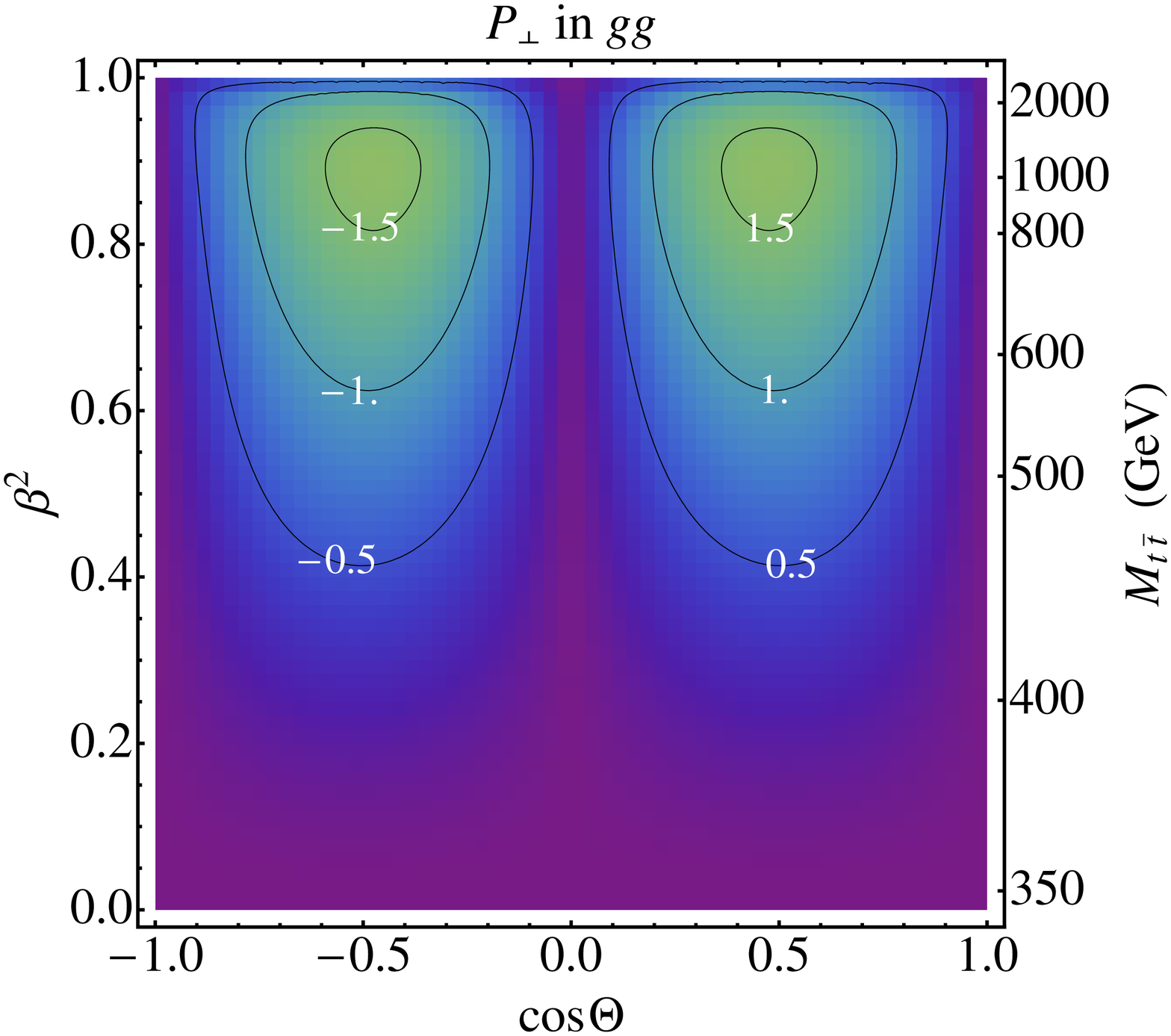}
\caption{\it QCD transverse polarization at $O(\alpha_s)$, in percent, for $q\bar q\to t\bar t$ (left) and $gg\to t\bar t$ (right) partonic subprocesses~\cite{Bernreuther:1995cx}, shown as a function of the top production angle and squared-velocity in the partonic CM frame.} 
\label{fig:yPol}
\end{center}
\end{figure}

At the Tevatron, there exists an obvious guess for the quark direction, namely the direction of the proton beam.  Using this direction to orient the production plane, and using the calculations of~\cite{Bernreuther:1995cx}, we estimate that the inclusive \Pperp\ at the Tevatron to $O(\alpha_s)$ is +1.1\%.  We illustrate the strength of the polarization over partonic production phase space in Fig.~\ref{fig:yPol}.  For completeness, we also illustrate the \Pperp\ from $gg$, though this plays no further role in our discussions.

At the LHC, the symmetric $pp$ initial state somewhat complicates the measurement of the $q\bar q \to t\bar t$ polarization.  However, we can apply the usual strategy for measuring forward-backward asymmetries at the LHC, namely by using the overall boost of the $t\bar t$ system relative to the lab-frame as a best-guess of the quark direction of motion.  This again integrates out any polarization from $gg$,\footnote{There is some subtlety here, as realistic acceptance cuts might leave over a net ``forward-backward asymmetry'' for CM production angles.  Since we define ``forward'' using the overall $t\bar t$ boost in the lab frame, tops emitted with $\Theta < \pi/2$ with respect to the forward beam in the partonic CM frame have a greater chance of being lost at high-$|\eta|$ than tops emitted with $\Theta > \pi/2$.  The situation is not quite so simple, as we must accept both the top and the antitop, and their production angles are highly anticorrelated.  However, even allowing for an acceptance bias, say against forward-emitted hadronic tops accompanying backward-emitted leptonic tops, the charge-flipped process in the same kinematic configuration from $gg\to t\bar t$ will have reversed polarization.  Any residual biases will therefore cancel out when we add together $t$ and $\bar t$ polarization measurements.} leaving over a net measurable $q\bar q$-initiated polarization of only +0.2\% at the 7~TeV and 8~TeV LHC.  The smallness of this number is a combination of the nontrivial measurement requirement and the dominance of the effectively unpolarized $gg$ process.

The polarization of top quarks must be inferred by studying their decay angles.  Because of the $V-A$ current structure of the decay, the lepton from the intermediate $W$ in semileptonic decay serves as a ``perfect'' spin analyzer up to very small higher-order corrections~\cite{Brandenburg:2002xr}.  Denoting the lepton's momentum direction in the top rest frame as $\hat\Omega(l)$, and integrating out the energies and orientations of the other decay products,
\beq
\frac{d^2\Gamma}{d\Omega(l)} \,\propto\, 1 + q_l\,\vec{P}\cdot\hat\Omega(l) \, ,
\eeq
where $q_l$ is the lepton's charge and $\vec{P} \equiv \langle 2\vec{S} \rangle$ is the top's (or antitop's) net polarization.  The relation holds for any lepton energy.  The most efficient way to probe for \Pperp\ is therefore to study the component of the lepton's momentum direction projected along the ``$y$-axis'' defined above:  $\cos\theta_y \equiv \hat y\cdot \hat\Omega(l)$.  Integrating out the corresponding azimuthal angle, 
\beq
\frac{d\Gamma}{d\cos\theta_y} \,\propto\, 1 + q_l \, P_\perp\cos\theta_y .
\eeq
In the absence of transverse polarization, this distribution is flat.  In the presence of transverse polarization, it acquires a linear bias, and an asymmetry
\beq
A_\perp^l = \frac{q_l\,P_\perp}{2}.
\eeq
Full event reconstruction with basic detector acceptance cuts can heavily resculpt the $\cos\theta_y$ distribution, but as we will see below the asymmetry largely persists.  So, in contrast to a forward-backward production asymmetry, we will study a ``left-right'' decay asymmetry, using the oriented production plane to define ``left'' and ``right.''

\section{Transverse Polarization and \AFB\ from Axigluon Models}
\label{sec:axigluon}

The smallness of \Pperp\ in QCD presents an interesting opportunity to probe for new physics effects.  This has been pointed out many times~\cite{Dharmaratna:1989jr,Dharmaratna2,Kane:1991bg,Bernreuther:1995cx,Godbole:2006tq}, however, there has been little discussion about what types of new physics would actually introduce an observable \Pperp.  The main ingredient that we require is a nontrivial complex phase between same-spin and opposite-spin amplitudes in the basis described above.  Motivated by the axigluon explanations of the Tevatron forward-backward asymmetry anomaly, we here restrict ourselves to understanding the \Pperp\ effects induced by these models.  As we will see shortly, the required phase is provided by the complex pole in the axigluon's propagator.  We reserve discussion of more general new physics scenarios to future work.

Axigluons are color-octet spin-one bosons that couple to quarks purely axially, in contrast to the purely vector couplings of ordinary gluons.  They can arise in models where the QCD $SU(3)$ symmetry arises from the spontaneous breakdown of a larger $SU(3)\times SU(3)$ gauge symmetry~\cite{Frampton:1987dn,Hall:1985wz}.  In more general models (e.g.,~\cite{Frampton:2009rk,Bai:2011ed,Tavares:2011zg}), the couplings to the different species of quarks can differ, and various mixtures of axial and vector couplings can arise.  Here we simply assume a common purely axial coupling for up and down quarks, and an independent purely axial coupling for top quarks.  We remain largely agnostic about the couplings to charm and strange, as these do not play a major role in \AFB\ or \Pperp\ due to their small PDF's, though the couplings should be small enough to evade dijet constraints from enhanced $s\bar s$ and $c\bar c$ production.  Similarly, we do not specify the coupling to right-handed bottom quarks.  (The coupling to left-handed bottoms is fixed by $SU(2)_L$ symmetry.)  The interaction Lagrangian is therefore
\beq
\Delta {\mathcal L}  \,=\, g_s A_\mu^a \, \Big( a_t \, \bar t \left[ T^a \gamma^\mu \gamma^5  \right]t   \: + \: a_q \sum_{q=u,d} \bar q \left[ T^a \gamma^\mu \gamma^5 \right]q   \: + \: (c,s,b\;{\rm couplings})  \Big) \, .
\eeq

As is by now well-appreciated, the interference between gluon exchange and axigluon exchange in $q\bar q \to t\bar t$ leads to a forward-backward asymmetry.  Expressed versus the $t\bar t$ invariant mass $\Mtt$, and integrated over production angles,
\beq
A_{FB}(\Mtt) = \frac{(3/2) a_q a_t  \beta (\Mtt^2 - M_A^2) \Mtt^4}{
\Big( (\Mtt^2 - M_A^2)^2 + \GA^2 M_A^2 \Big) \Big(\Mtt^2 + 
    2 m_t^2\Big)   \,+\,  a_q^2 a_t^2 \beta^2 \Mtt^6} \, ,
\eeq
where $M_A$ is the axigluon mass, $\GA$ is its width, and $\beta = \sqrt{1-4m_t^2/\Mtt^2}$ is the top or antitop velocity.  Introducing vector couplings yields a more complicated expression, but the overall effect is always proportional to $a_q a_t$.  The distribution of \AFB\ versus $\Mtt$ has been measured by CDF~\cite{Aaltonen:2012it} and D0~\cite{Abazov:2011rq}, as well as the inclusive \AFB\ and purely leptonic forward-backward asymmetries~\cite{CDF:AFB2011dilep,Abazov:2011rq,Abazov:2012bfa}.  All of these measurements show some degree of tension with the Standard Model, usually at the 2--3$\sigma$ level.  In particular, the inclusive measurements, corrected for detector effects, exceed the QCD prediction of approximately 6.6\%~\cite{Aaltonen:2012it} by 9.8$\pm$4.9\% (CDF) or 13.0$\pm$6.8\% (D0).  With appropriately chosen parameters, an axigluon resonance can make up this difference.  Because the predicted QCD asymmetry and the anomaly are both much smaller than one, it is adequate to simply add the axigluon's tree-level asymmetry contribution.

Notably, the axigluon's \AFB\ goes through a zero at $\Mtt = M_A$, where the gluon and axigluon propagators are 90-degrees out of phase and the rate interference vanishes.  Because of this zero, the relative signs required of $a_q$ and $a_t$ depend on the axigluon's mass.  For ``light'' axigluons, near or below $t\bar t$ production threshold, the couplings should be relatively positive, possibly even flavor-universal.  For ``heavy'' axigluons, above 450--500~GeV, the couplings must be relatively negative, requiring a flavor bias that discriminates between top quarks and light quarks.

Both situations have been proposed to resolve the \AFB\ anomaly.  However, both also face tight constraints from the Tevatron and LHC, which have conducted searches for $t\bar t$ resonances and other deviations of the differential mass spectrum~\cite{Aaltonen:2009iz,CDFresonance,Abazov:2011gv,ATLAScombinedResonance,Aad:2012raa,Aad:2012hg,Chatrchyan:2012cx,CMSboosted,CMS:2012qka}, dijet resonances and contact interactions~\cite{CMS:2012yf,Chatrchyan:2013muj,ATLASdijetResonance,ATLAS:2012pu}, pair production of new particles decaying to jets~\cite{Chatrchyan:2013izb,ATLAS:2012ds,CDF4j}, as well as the \AFB-related forward-central charge asymmetry ($A_C$) at the LHC~\cite{Chatrchyan:2012cxa,ATLAS:2012an,ATLASchargeAsymDilep}.\footnote{See also~\cite{Haisch:2011up,Gresham:2012kv} for discussions of electroweak constraints, which are mainly relevant for light axigluons below $t\bar t$ production threshold.  Limits from four-top production via axigluon pairs can also be relevant for heavier axigluons, and are discussed in~\cite{AguilarSaavedra:2011ck}.}  For light axigluons, the most promising option to escape detection is to assume a large $\GA/M_A$ (a few 10's of \%), which significantly weakens discrimination from continuum backgrounds.  The large width might arise from new colored decay channels, which themselves are subject to stringent constraints but can escape detection if they result in high-multiplicity multijet final states~\cite{Tavares:2011zg,Gross:2012bz}.  Heavier axigluons also remain a viable possibility, with some qualifications.  These are naively somewhat better-hidden due to their inefficient on-peak production at the Tevatron and the large $gg\to t\bar t$ background at the LHC, but they also require much larger couplings to provide the Tevatron \AFB.  To escape detection in both $t\bar t$ and dijet searches, such axigluons would ideally also be very broad and/or have attenuated couplings to light quarks.  These two options actually go hand-in-hand, as couplings biased highly in favor of tops also cause the width to become large.  For example, taking the top (and $b_L$) couplings on the high side while staying marginally perturbative, $a_t \simeq 5$, we expect $\GA/M_A \simeq 0.5$.\footnote{Of course, with such large widths there is little guarantee that the full axigluon propagator looks like that of a free particle with a complex pole mass.  However, for want of any more compelling model of the full behavior, we continue to use this functional form as our ansatz for all widths.}  Alternatively, the axigluon may simply be too heavy to produce with appreciable on-peak rate even at the LHC, so that it is practically felt as a contact operator~\cite{Delaunay:2011gv}.  This results in a significant growth of the $t\bar t$ differential cross section at high energy, as well as $A_C$, and is currently in modest tension with the LHC measurements~\cite{Delaunay:2012kf}.  However, keeping this as an open option, it is also likely broad without the need for additional decay channels, because of the large couplings required by \AFB\ ($a_q a_t \simeq (M_A/{\rm TeV})^2$).

Broad axigluons automatically induce a transverse top polarization, again via interference with QCD and again in direct proportion to the axial couplings.  Assuming, as we are, vanishing vector couplings, we get
\beq
P_\perp(\Mtt) \,=\,  \frac{(-3\pi/4) a_q a_t  \beta  m_t \GA M_A^2 \Mtt^3}{
\Big( (\Mtt^2 - M_A^2)^2 + \GA^2 M_A^2 \Big)  \Big(\Mtt^2 + 
    2 m_t^2\Big)   \,+\,  a_q^2 a_t^2 \beta^2 \Mtt^6} \, . 
\eeq
The strength of \Pperp\ is necessarily directly proportional to $\GA$, as the gluon and axigluon exchange diagrams become relatively real in the limit $\GA \to 0$.  As per our discussions leading to Eq.~\ref{eq:PperpQCD}, the numerator also contains a factor of the velocity $\beta$ and, because axigluons also respect chirality, a factor of $m_t$.  While the QCD+axigluon theory is CP- and T-symmetric, and therefore also subject to naive time-reversal invariance and vanishing of \Pperp\ at tree-level, the width of the axigluon is formally a loop effect.\footnote{In~\cite{Baumgart:2012ay}, we pointed out a similar effect of this form, in a parity-violating, CP-conserving spin correlation $\vev{S_x \bar S_y} = \vev{S_y \bar S_x} \ne 0$.  This correlation violates parity because under that transformation (when combined with 180-degree rotation) $S_x$ flips sign but not $\bar S_y$.  The correlation also violates naive time-reversal, which (again times 180-degree rotation) flips $\bar S_y$ but not $S_x$.  The effect is maximized when the light quark couplings are pure vector and the top quark couplings are pure axial.}  With exactly vanishing vector couplings, the axigluon also separately respects C- and P-symmetries, and hence \Pperp\ would be the {\it only} net polarization effect allowed.  Nonetheless, the presence of the axigluons will generally alter the $t\bar t$ spin correlations, a point which has been emphasized in~\cite{Baumgart:2011wk,Krohn:2011tw,Fajfer:2012si}, and to which we will briefly return below.

For a given $\Mtt$, the ratio between \Pperp\ and \AFB\ is very simple\footnote{If we turn on vector couplings, the denominator picks up an extra term $2v_q v_t \Mtt^3$, from the shift to \AFB.}:
\beq
\frac{P_\perp(\Mtt)}{A_{FB}(\Mtt)} \,=\, \frac{(-\pi/2) m_t \GA M_A}{\Mtt(\Mtt^2 - M_A^2)} \, .
\eeq
In the case of a heavy axigluon, we can already infer an approximate relationship between the inclusive \Pperp\ and \AFB\ at the Tevatron by assuming $M_A^2 \gg \Mtt^2$ and $\Mtt \sim 2m_t$:  $P_\perp / A_{FB} \sim (\pi/4)(\GA/M_A)$.  The resulting leptonic \Aperpl\ is then half again this size.  Given that \AFB\ anomaly is itself on the borderline of significance, and $\GA/M_A < 1$ in any ``reasonable'' model, it seems unlikely that the Tevatron will be capable of probing the \Pperp\ induced by heavy axigluons, leaving these for investigation at the LHC.

\begin{figure}[tp]
\begin{center}
\epsfxsize=0.49\textwidth\epsfbox{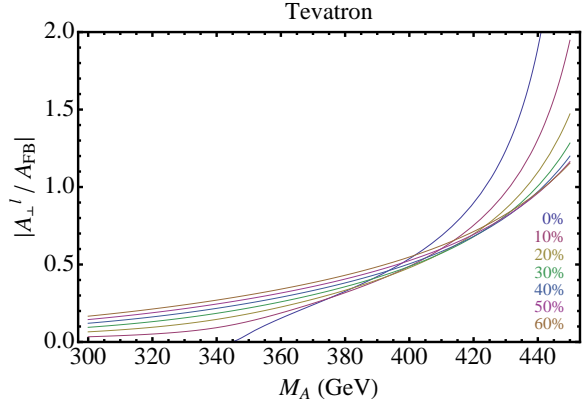}
\caption{\it Ratio between the inclusive lepton production plane reflection asymmetry (\Aperpl) and the inclusive $t\bar t$ forward-backward asymmetry (\AFB) at the Tevatron, induced by an axigluon resonance with vanishing vector couplings.  Different colored lines indicate different $\GA/M_A$ values.  The ratio is independent of the strength of the axigluon's couplings.} 
\label{fig:AperpOverAFB}
\end{center}
\end{figure}

However, if the axigluon pole is within the energy range of the Tevatron, we see that \Pperp\ can be much larger, relatively speaking.  In Fig.~\ref{fig:AperpOverAFB}, we show the ratio between inclusive \Aperpl\ and \AFB\ at the Tevatron, as a function of $M_A$ for several $\GA/M_A$ assumptions.\footnote{For completeness, we include predictions for narrow axigluons.  Indeed, even the limit $\GA/M_A \to 0$ produces an appreciable integrated \Aperpl relative to \AFB.  However, these likely do not serve as realistic model points, as the coupling strengths required to reproduce the measured \AFB\ would lead to a significant modification to the total $t\bar t$ cross section and a large resonance peak in the $t\bar t$ mass spectrum.}  Note that for purely axial couplings this ratio {\it only} depends on $M_A$ and $\GA/M_A$, since the total transverse-polarized cross section and total $\sigma_F - \sigma_B$ are both proportional to $a_q a_t$.  The total cross section (which contains factors $\propto a_q^2 a_t^2$) cancels out.  We can see in Fig.~\ref{fig:AperpOverAFB} that the \Aperpl$/$\AFB\ increases steadily with the axigluon's mass within the plotted range, but does not have a simple dependence on the axigluon's width.  For axigluons very close to top pair threshold or even below threshold, the width has only a minor impact on \AFB, whereas it has a quite direct effect on \Aperpl.  The ratio therefore increases with increasing $\GA/M_A$.  For axigluons near 450--500~GeV, the contributions from below peak and above peak approximately cancel in \AFB\ but not in \Aperpl.  The former then becomes the stronger function of the axigluon width, and $A_\perp^l/A_{FB}$ {\it decreases} with increasing $\GA/M_A$.  For a broad swath of axigluon masses in between, the two tendencies approximately balance, and $A_\perp^l/A_{FB}$ is actually fairly stable versus the axigluon width.  Ratios of $O$(0.3--1) are typical in this range.  While these numbers indicate that \Aperpl\ is generally somewhat smaller than \AFB\ for light axigluons at the Tevatron, we point out that the former is maximized for central tops, whereas the latter is maximized for forward tops.  Detector acceptance cuts, which favor more central tops, therefore tend to accentuate \Aperpl\ and suppress \AFB.  The construction of \Aperpl\ at detector-level is also simpler in principle than \AFB.  We will discuss Tevatron measurement prospects in much more detail in the next section.

At the LHC, the effects of \Aperpl\ are inevitably attenuated due to the large $gg\to t\bar t$ background and the fact that the initial quark (versus antiquark) direction must be guessed based on the partonic $t\bar t$ system's longitudinal boost.  On the other hand, the LHC benefits from dramatically larger overall statistics and much easier access to larger $t\bar t$ invariant masses.  In particular, it is possible to construct fairly specialized cuts to target interesting regions of production phase space, where the Tevatron would quickly run out of events.  We will also explore these points in the next section.

To establish the detailed phenomenology at the Tevatron and LHC, we now focus on three choices for axigluon mass and width.  As an example of a ``light'' resonance that can lead to observable \Pperp\ effects at the Tevatron, we set the mass and width to 420~GeV and 20\%, respectively.  These parameters correspond to one of the models studied in~\cite{Tavares:2011zg}.  As an example of a ``heavy'' resonance, which is nonetheless light enough to be within range of the LHC, we set the mass and width to 800~GeV and 50\%, respectively.  Finally, we also consider the 4-quark contact-interaction limit, taking $M_A \to \infty$ but keeping $\GA/M_A$ fixed to 50\%.

To determine the appropriate couplings for these three models, we combine the CDF and D0 inclusive asymmetry measurements according to their inverse-squared errors, and determine the averaged excess relative to the SM expectation.  Summing the measurement errors and theoretical errors in quadrature (assuming 2\% for the latter) yields an \AFB\ excess of $10.8\pm4.2\%$.  We allow our model couplings to span the $\pm1\sigma$ range.  For the light axigluon, this corresponds to $a_q a_t = [0.11,0.29]$.  For the heavy axigluon, it corresponds to $a_q a_t = -[0.34,0.78]$.  For the contact interaction, we require $a_q a_t / M_A^2 = -[0.56,1.23]$~TeV$^{-2}$.  (Note that the finite width generally attenuates the contact interaction's \AFB\ by $1+(\GA/M_A)^2$, relative to the usual narrow-width assumption.)

\begin{figure}[tp]
\begin{center}
\epsfxsize=0.49\textwidth\epsfbox{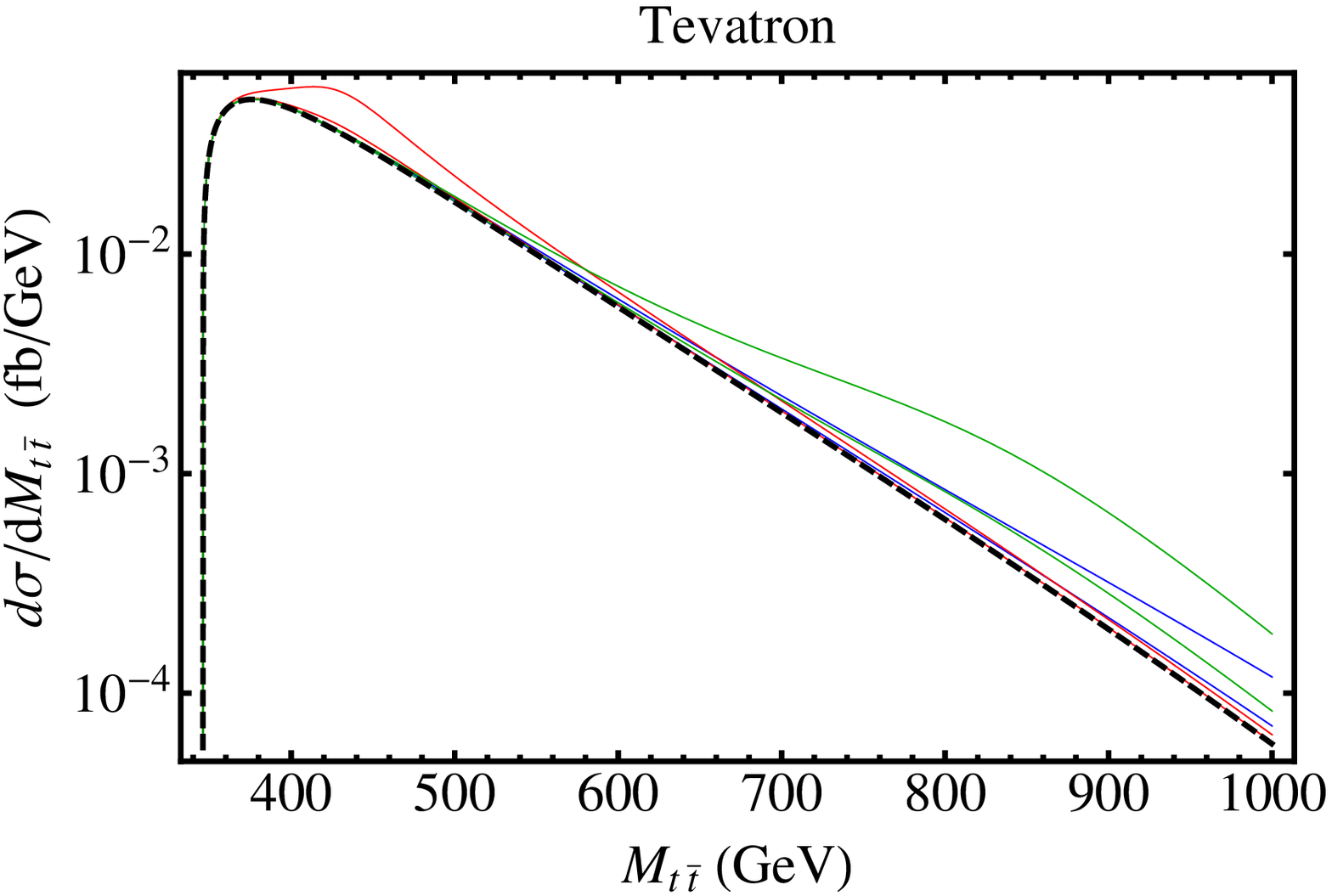}
\epsfxsize=0.49\textwidth\epsfbox{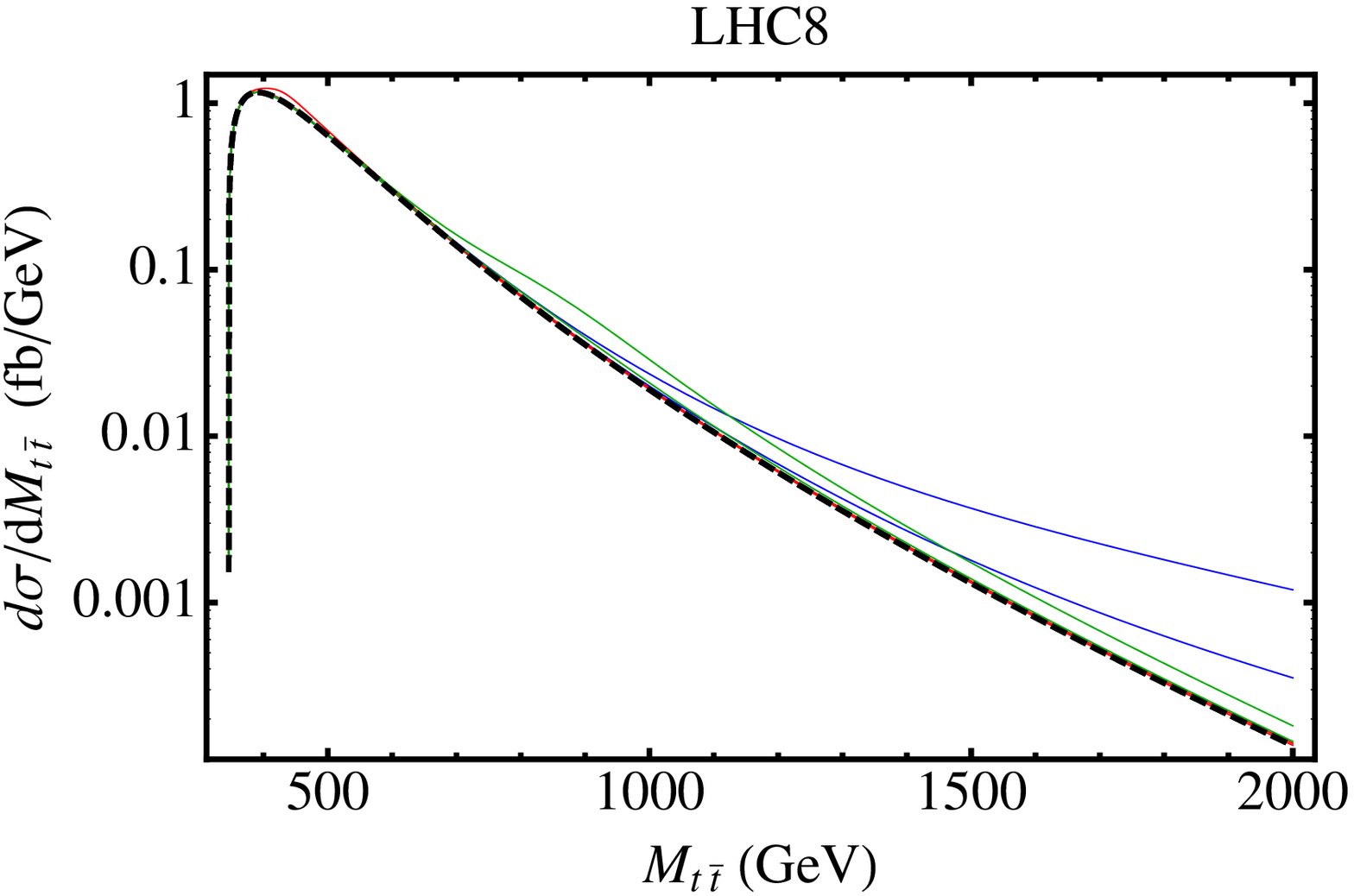}
\caption{\it The $t\bar t$ differential cross section versus $\Mtt$ at the Tevatron (left) and 8~TeV LHC (right).  The SM prediction is the dashed black line.  The predictions with a 420~GeV axigluon with 20\% width (red), an 800~GeV axigluon with 50\% width (green), and an axial contact interaction with 50\% width (blue) are shown for coupling strengths that yield the inclusive Tevatron \AFB\ excess $\pm 1\sigma$.} 
\label{fig:dsigModels}
\end{center}
\end{figure}

\begin{figure}[tp]
\begin{center}
\epsfxsize=0.49\textwidth\epsfbox{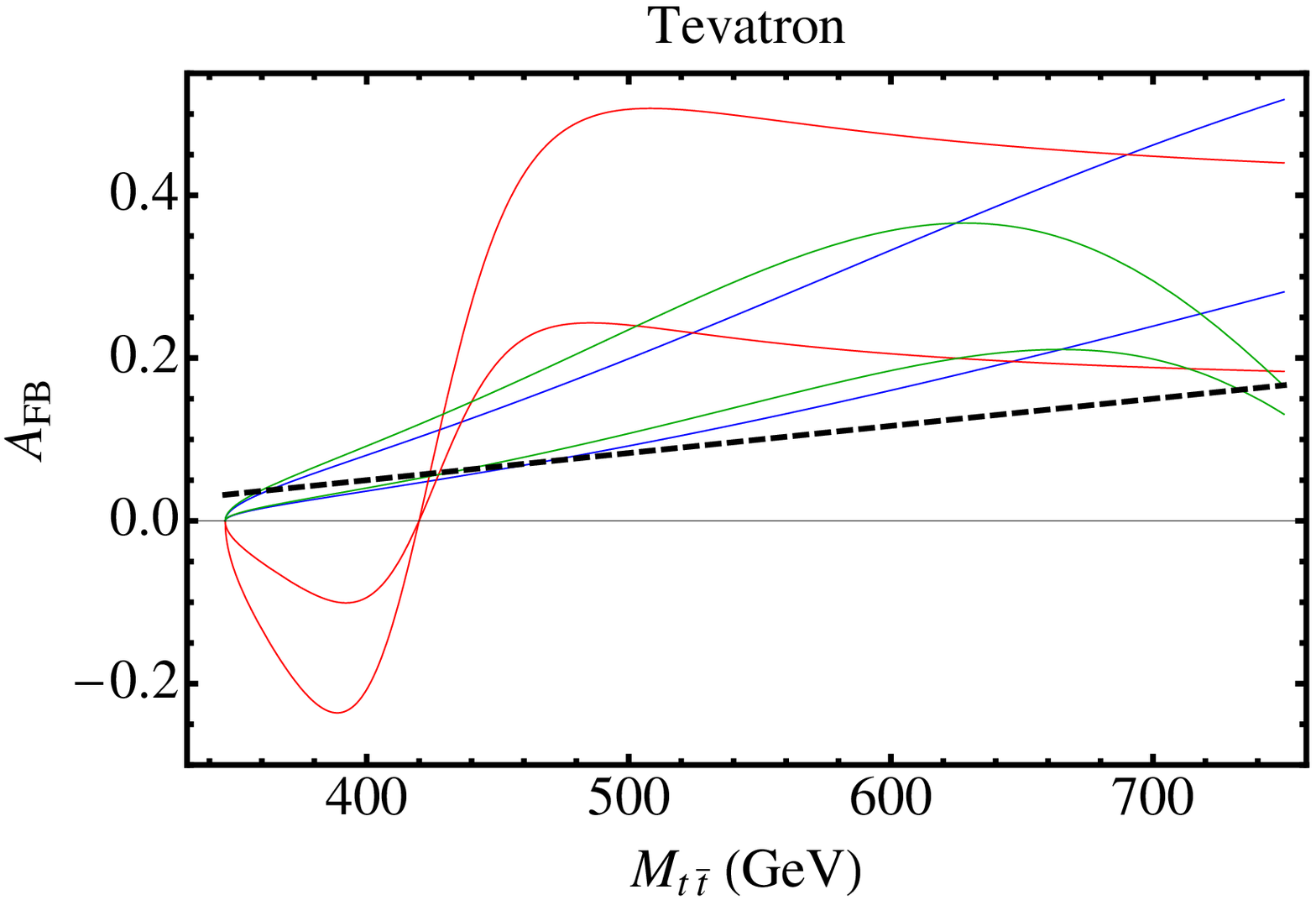}
\epsfxsize=0.49\textwidth\epsfbox{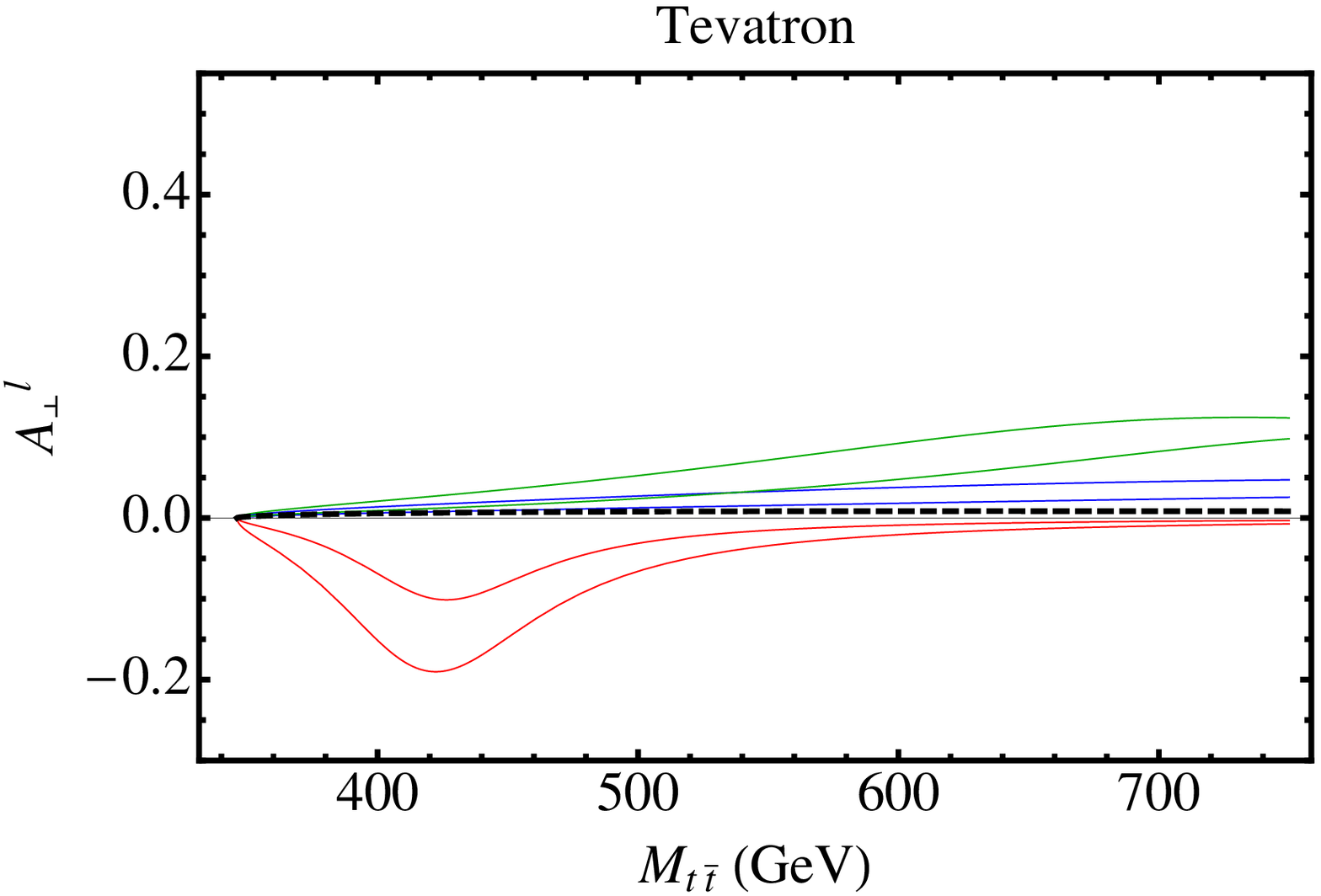}
\caption{\it Induced $t\bar t$ forward-backward asymmetry (\AFB, left) and lepton production plane reflection asymmetry (\Aperpl, right) as a function of $\Mtt$ at the Tevatron.  The SM prediction is the dashed black line.  The non-SM contributions from a 420~GeV axigluon with 20\% width (red), an 800~GeV axigluon with 50\% width (green), and an axial contact interaction with 50\% width (blue) are shown for coupling strengths that yield the inclusive Tevatron \AFB\ excess $\pm 1\sigma$.} 
\label{fig:TevatronModels}
\end{center}
\end{figure}

\begin{figure}[tp]
\begin{center}
\epsfxsize=0.49\textwidth\epsfbox{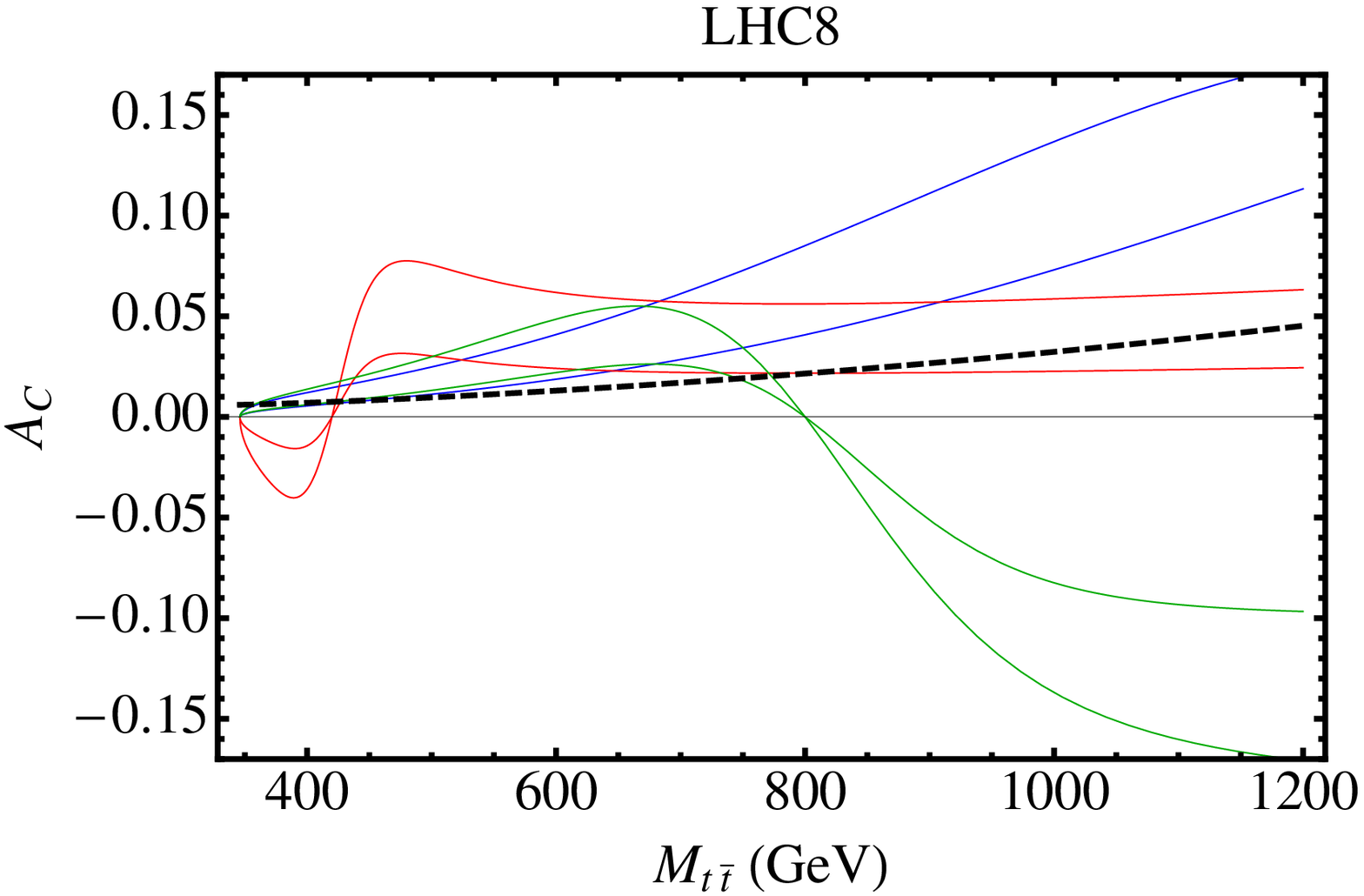} \\ \vspace{0.3in}
\epsfxsize=0.49\textwidth\epsfbox{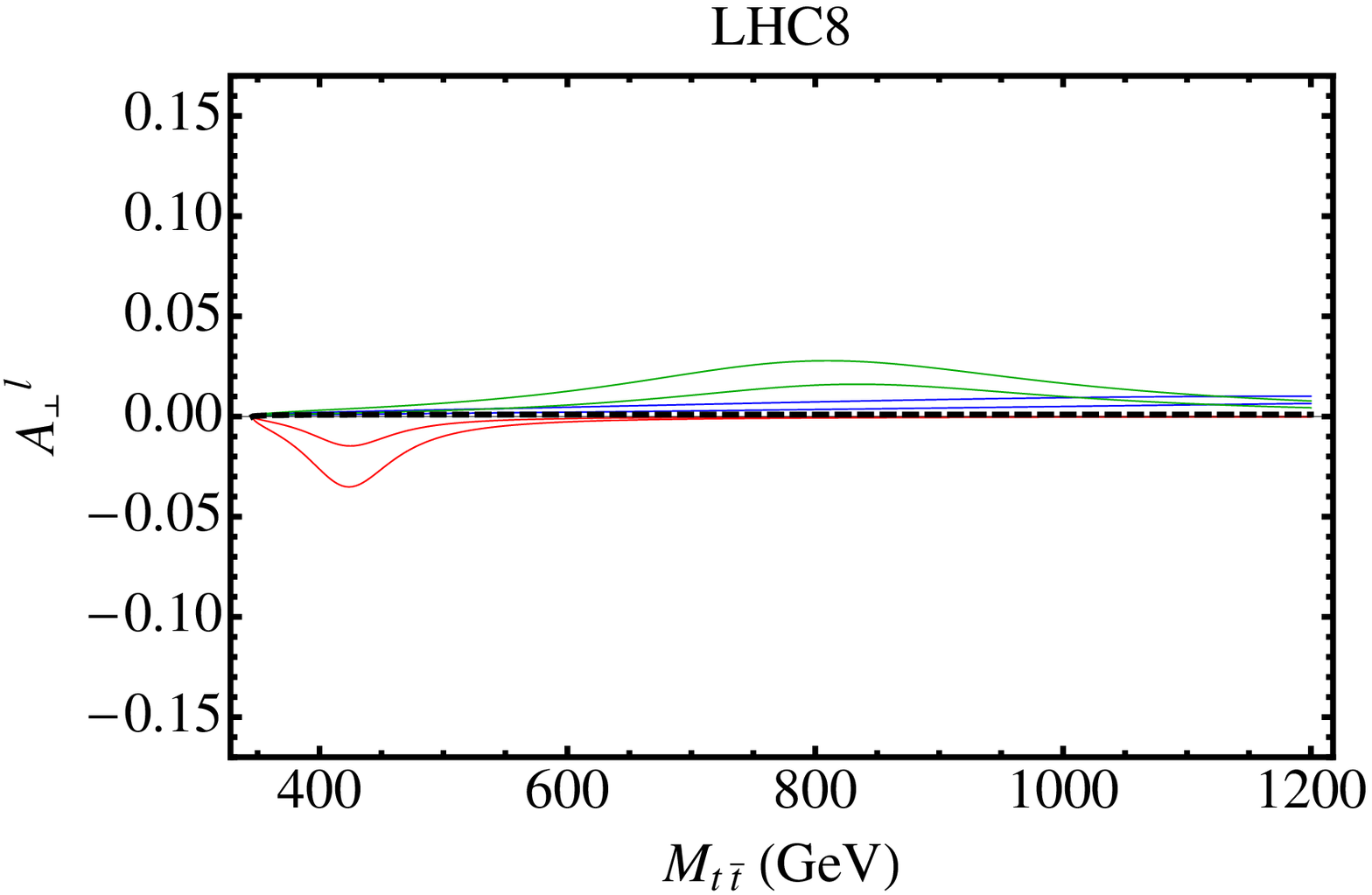}
\epsfxsize=0.49\textwidth\epsfbox{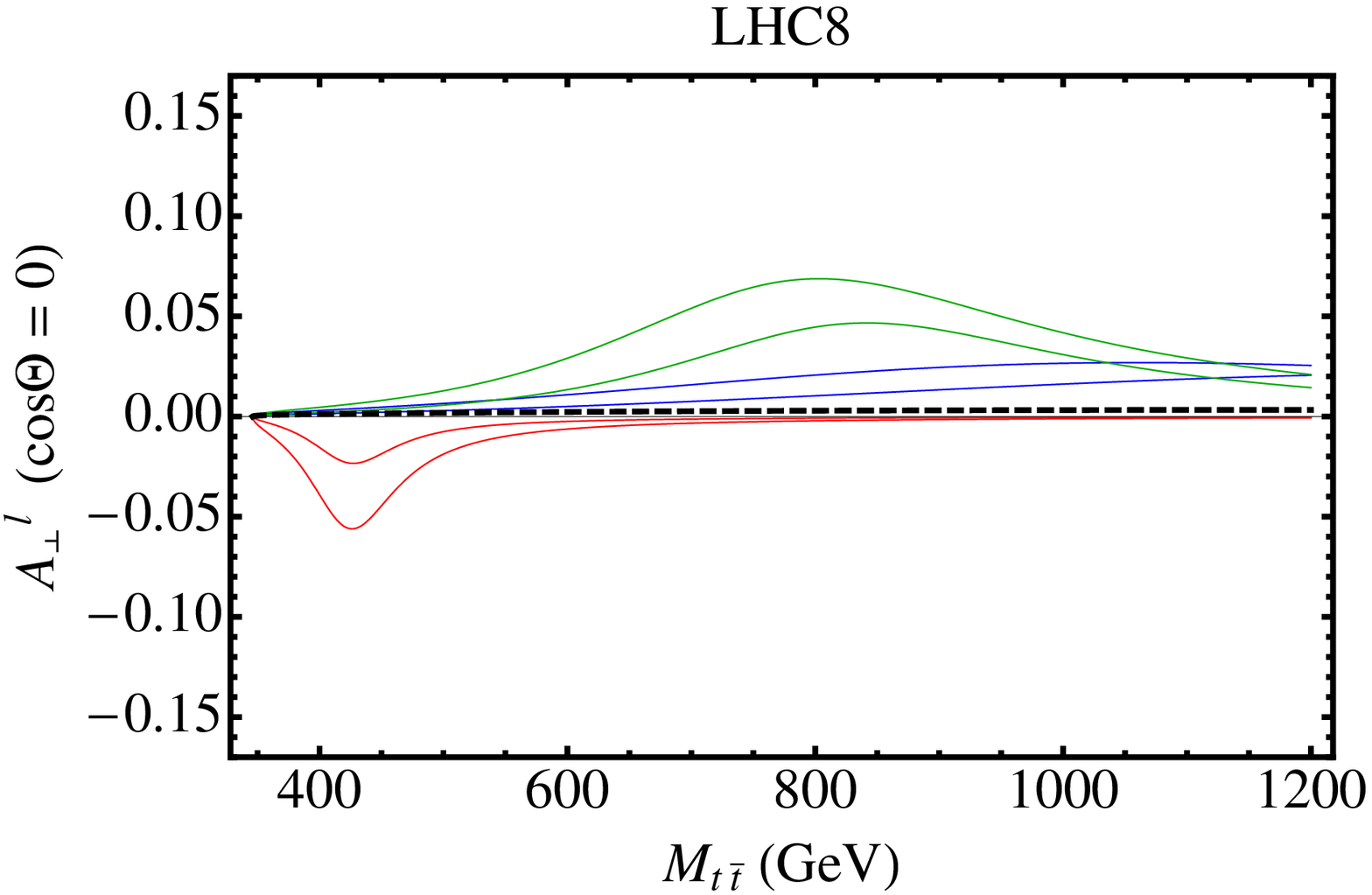}
\caption{\it Induced $t\bar t$ forward-central charge asymmetry ($A_C$, top), lepton production plane reflection asymmetry (\Aperpl, bottom left), and \Aperpl\ at central partonic CM production angles (bottom right) as a function of $\Mtt$ at the 8~TeV LHC.  The SM prediction is the dashed black line.  The non-SM contributions from a 420~GeV axigluon with 20\% width (red), an 800~GeV axigluon with 50\% width (green), and an axial contact interaction with 50\% width (blue) are shown for coupling strengths that yield the inclusive Tevatron \AFB\ excess $\pm 1\sigma$.} 
\label{fig:LHCModels}
\end{center}
\end{figure}

We display the models' effects on the $t\bar t$ invariant mass spectra at the Tevatron and LHC in Fig.~\ref{fig:dsigModels}.  In Fig.~\ref{fig:TevatronModels}, we show the induced differential $t\bar t$ forward-backward asymmetry and \Aperpl at the Tevatron.  In Fig.~\ref{fig:LHCModels}, we show the corresponding $A_C$ and \Aperpl\ at the LHC.  There, we also include a plot of \Aperpl\ for central CM-frame production angles, where the size of the $q\bar q$ asymmetry is enhanced and the relative $gg$ contamination is reduced.  The $A_C$ and \Aperpl\ measurements at the LHC both rely on the overall $t\bar t$ system boost relative to the lab to define a best-guess $q\bar q$ initial-state orientation.\footnote{The forward-central charge asymmetry is usually defined directly as the asymmetry between $N(\Delta|y|>0)$ and $N(\Delta|y|<0)$, with $\Delta|y| \equiv |y_t| - |y_{\bar t}|$.  This is exactly the same as the CM-frame forward-backward asymmetry, defining ``forward'' using the CM boost direction in the lab.}  There may also be some benefit to measuring the ratio, \Aperpl$/ A_C$ (binned in $M_{t \bar t}$ and/or the $t \bar t$ system rapidity).  Near a resonance, this quantity can become very large.  Additionally, the dependence on the gluon PDF divides out, removing an important source of systematic uncertainty.

\begin{figure}[tp]
\begin{center}
\epsfxsize=0.49\textwidth\epsfbox{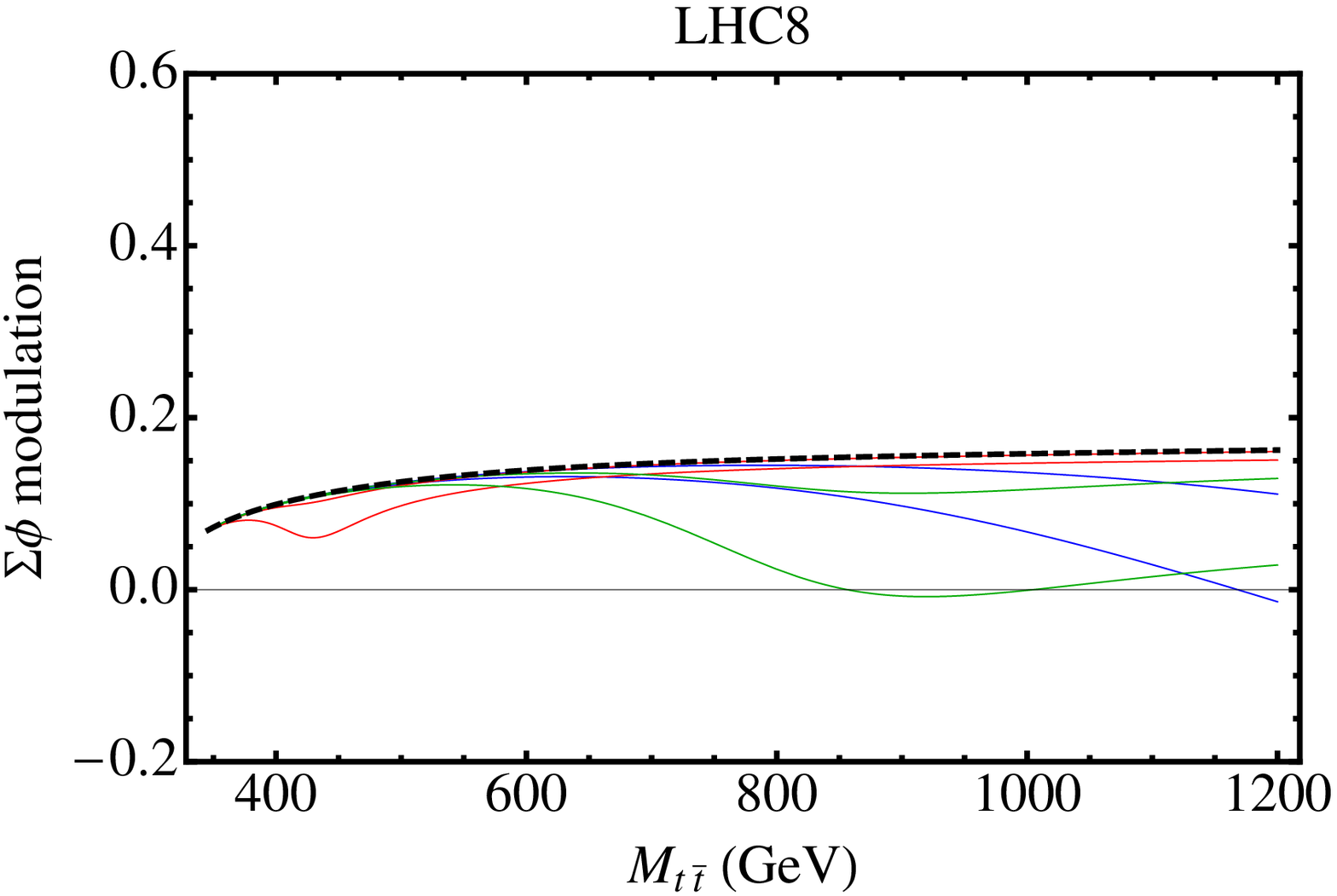}
\epsfxsize=0.49\textwidth\epsfbox{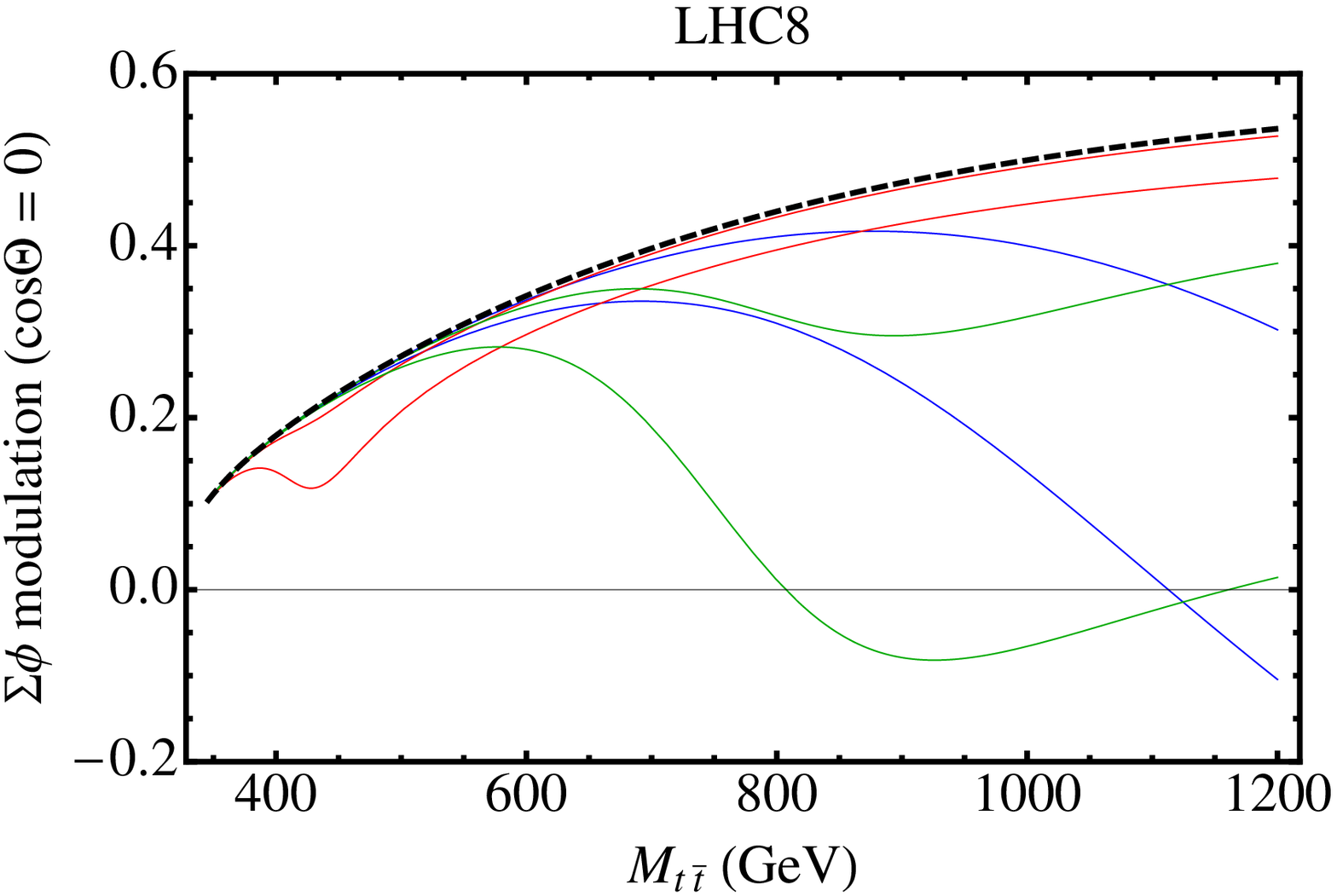}
\caption{\it Fractional modulation amplitude in the summed top and antitop azimuthal decay angles (``$\Sigma\phi$'') as a function of $\Mtt$ at the 8~TeV LHC, inclusive over partonic CM production angles (left) and at central angles (right).  Spin analyzing powers have not been included.  The SM prediction is the dashed black line.  The predictions with a 420~GeV axigluon with 20\% width (red), an 800~GeV axigluon with 50\% width (green), and an axial contact interaction with 50\% width (blue) are shown for coupling strengths that yield the inclusive Tevatron \AFB\ excess $\pm 1\sigma$.} 
\label{fig:phiSum}
\end{center}
\end{figure}

In~\cite{Baumgart:2011wk}, we also studied the effects of axigluon (and axial $Z'$) resonances on $t\bar t$ spin {\it correlations}.  The dominant effect occurs in a somewhat non-obvious sum of the two tops' azimuthal decay angles about their common production axis, inducing a cosine-wave modulation.  The sum can be formed between any single top decay product and antitop decay product, such as two leptons or a lepton and a jet.  In Fig.~\ref{fig:phiSum}, we show how our current set of models would affect this modulation amplitude at the LHC.  Unlike the transverse polarization, the pure QCD contribution is already sizable, but the axigluons nonetheless cause significant distortions.  We return to this point when we discuss measurements in the next section.

\begin{table}[tp]
 \centering
\begin{tabular}{l|r|r|r|r|r|r}
    Model        & \; $\GA/M_A$ \;  &  $a_q a_t$ \;                 &  \; Tevatron \AFB\ \; & \; Tevatron \Aperpl\ \; & \; LHC $A_C$ \; & \; LHC \Aperpl\ \\ \hline \hline
420 GeV, central &  20\%            &  0.18                         &  10.8\%               &  -7.5\%                 & 2.3\%           &  -1.1\% \\ 
420 GeV, lower   &  20\%            &  0.11                         &   6.6\%               &  -4.8\%                 & 1.4\%           &  -0.7\% \\ \hline
800 GeV, central &  50\%            & -0.55                         &  10.8\%               &   3.0\%                 & 1.5\%           &   0.6\% \\
800 GeV, lower   &  50\%            & -0.34                         &   6.6\%               &   1.9\%                 & 0.9\%           &   0.4\% \\ \hline
contact, central &  50\%            & -0.89$(\frac{M_A}{\rm TeV})^2$ &  10.8\%              &   1.5\%                 & 2.4\%           &    0.3\% \\
contact, lower   &  50\%            & -0.56$(\frac{M_A}{\rm TeV})^2$ &   6.6\%              &   0.9\%                 & 1.5\%           &    0.2\% \\
\end{tabular}
\caption{\it Benchmark models for our subsequent analyses and their parton-level contributions to the inclusive Tevatron and LHC asymmetries.  ``Central'' and ``lower'' refer to whether the model reproduces the central value of the measured Tevatron $A_{FB}$ excess or its $-1\sigma$ value.  The corresponding models that reproduce the $+1\sigma$ value are excluded and henceforth not considered.}
\label{tab:models}
\end{table}

The +$1\sigma$ ends of our models' coupling ranges are in conflict with one or more $t\bar t$ measurements, including the unfolded differential cross sections at the Tevatron~\cite{Aaltonen:2009iz} and LHC~\cite{Aad:2012hg,CMS:2012qka}, the LHC $A_C$ measurements~\cite{Chatrchyan:2012cxa,ATLAS:2012an}, and also the mass-binned \AFB\ at CDF for the light axigluon~\cite{Aaltonen:2012it}.  However, the $-1\sigma$ couplings would have easily escaped detection in all cases, largely due to the fact that the effect on the mass spectrum is quartic in the couplings, versus quadratic for \AFB.  Values that reproduce the central \AFB\ tend to be borderline allowed.\footnote{For the light axigluon, tension with data is strongest for the Tevatron differential \AFB\ measurement.  The heavy axigluon and contact operator scenarios are most constrained by the high-mass bins in differential cross-section for CDF and ATLAS, respectively.  In all of these cases, the model is at or just beyond 95\% exclusion.  For simplicity, we thus keep the ``central'' value for couplings as the upper limit we consider.}  Therefore, for the remainder of our study, we will only consider couplings spanning from $-1\sigma$ to central.  Starting from this reduced range, we will see whether transverse polarization can have a role to play in diagnosing the source of the \AFB\ anomaly.  We list our set of six benchmark models in Tab.~\ref{tab:models}.


\section{Measurement Prospects}
\label{sec:measurement}

In this section, we will estimate what might be possible for realistic measurements of \Pperp\ via \Aperpl.  We consider measurements at both the Tevatron and LHC, as well as in the $l$+jets and dileptonic channels.  Detailed descriptions of our signal and background simulations and basic event reconstructions can be found in Appendix~\ref{sec:details}.

Measuring \Aperpl\ requires us to define an oriented production plane for each $t\bar t$ candidate event and check how often the lepton appears on either side.  We construct the oriented normal by crossing together the initial quark vector and the transverse momentum vector of one of the tops.  A major advantage of this construction is that we do not strictly need a fully global picture of the $t\bar t$ system.  At the Tevatron, where the orientation of $q\bar q$ system is essentially known a priori, it is adequate to just measure the $\vec{p}_T$ vector of a leptonic top.  We also re-emphasize that, unlike \AFB, \Aperpl\ is maximized for central production, which puts the tops in the region of largest acceptance for realistic detectors, and also with the smallest $gg\to t\bar t$ background at the LHC.  The measurement of \Aperpl\ is, in more ways than one, ``orthogonal'' to the measurement of \AFB\ or $A_C$.

Estimating the transverse production axis is actually very straightforward.  For example, in both $l$+jets and dileptonic events, the transverse thrust axis usually provides a good approximation.  The real challenge, though, is to orient this axis.  Unless the tops are moving with appreciable $p_T$, the reconstruction of one or both tops poses a nontrivial combinatorial problem.  Because the mass scales of the two decays are quite similar, it is easy to mistakenly assign a given jet or lepton to the wrong side of the event.  One of the main goals of this section is to demonstrate that the combinatorial problem can be largely overcome.

At the LHC, which is a symmetric $pp$ collider, we face the additional challenge of measuring the orientation of the initial light quark and antiquark.  While this is not possible to do unambiguously on an event-by-event basis, we can exploit the fact that the quark PDF's are harder than the antiquark PDF's.  The longitudinal boost of the $t\bar t$ system with respect to the lab, $\Ytt$, is then correlated with the quark direction.  The correlation becomes stronger with larger $\Ytt$.  While the asymmetry is inevitably washed-out to some degree, the enormous statistics available at the LHC more than compensate.

In what follows, we will explore several measurement strategies, with varying degrees success.  Generally, we find that the $l$+jets channel is the most promising at both machines, owing to the larger statistics and better control over event kinematics.  However, the effect might also be visible in the dileptonic channel, where each event provides us with two \Aperpl's to measure.  For our 420~GeV axigluon models, \Aperpl\ at the Tevatron can be larger than 6\% at reconstruction-level in $l$+jets, allowing nearly 3$\sigma$ statistical sensitivity with the full data set.  At the LHC, all of our benchmark models are accessible to at least the 2$\sigma$ level, and the 800~GeV model is visible at 6$\sigma$.

In deriving all of our results below, we neglect the intrinsic \Aperpl\ from QCD, which we expect to be much smaller than our signals.

\subsection{Tevatron}
\label{sec:Tevatron}

We start with our $l$+jets analysis.  Perhaps the most straightforward approach is to perform a kinematic $\chi^2$ minimization over all possible assignments of jets to the leptonic and hadronic top and the two possible neutrino $p_z$ solutions.  CDF and D0 use rather sophisticated multidimensional $\chi^2$ functions that also allow the individual measured momenta to vary within experimental errors.  We use a much simpler method, which keeps the kinematics fixed and iterates over partitionings amongst the four hardest jets into $l\nu j$ and $jjj$ subsystems, picking the one that minimizes $(m(l\nu j) - m_t)^2 + (m(jjj) - m_t)^2$.  If the event contains one tagged $b$-jet, it must be used in the reconstruction of one of the tops.  If it contains two or more tagged $b$-jets, both of the tops must contain a $b$-tagged jet.  If no real neutrino solution exists, we reduce the magnitude of \met\ to obtain one.  To get some sense of whether this reconstruction furnishes a reasonable approximation to those of the real Tevatron experiments, we have cross-checked against CDF's reconstruction of \AFB.  We obtain a good reproduction of the $\Delta y$ response matrix (Fig.~9 in~\cite{Aaltonen:2012it}), and observe a realistic $O(0.5)$ dilution factor between the induced parton-level asymmetry and reconstruction-level asymmetry from heavy axigluons.

\begin{table}[tp]
 \centering
\begin{tabular}{l||r|r|r||r}
Tevatron, 9.4 fb$^{-1}$ &                      &                        &                                  &            \\ 
$l$+jets              & \; $\ge 4j$ Global \; & \; $\ge3j$ $t_l$ \;     & \; $\ge 3j$ $t_l$/$t_h$ Agree \; & \; $\ge3j$ Perfect \;   \\ \hline \hline
\# $t\bar t$ events   &  2200                 &  3550                   &  2440                           &  3550      \\
$S/B$                 &  4.1                  &  1.6                    &  1.6                            &            \\ \hline \hline
420 GeV, central      &  4.7\% ($2.0\sigma$)  & \;  4.7\% ($2.2\sigma$) &  6.7\% ($2.6\sigma$)            &  8.7\%     \\ 
420 GeV, lower        &  2.9\% ($1.2\sigma$)  & \;  3.0\% ($1.4\sigma$) &  4.2\% ($1.6\sigma$)            &  5.4\%     \\ \hline
800 GeV, central      & -3.1\% ($1.3\sigma$)  & \; -3.1\% ($1.4\sigma$) & -4.0\% ($1.5\sigma$)            & -4.5\%     \\
800 GeV, lower        & -2.0\% ($0.8\sigma$)  & \; -2.0\% ($0.9\sigma$) & -2.5\% ($1.0\sigma$)            & -3.0\%     \\ \hline
contact, central      & -1.3\% ($0.5\sigma$)  & \; -1.4\% ($0.6\sigma$) & -1.7\% ($0.7\sigma$)            & -2.3\%     \\
contact, lower        & -0.8\% ($0.3\sigma$)  & \; -0.9\% ($0.4\sigma$) & -1.1\% ($0.4\sigma$)            & -1.6\%     \\
\end{tabular}
\caption{\it Tevatron (9.4~fb$^{-1}$) reconstruction-level \Aperpl\ in $l$+jets $t\bar t$ events for our baseline models using our different reconstruction strategies, and with a perfect leptonic top reconstruction for reference.  We also indicate the statistical significance, assuming a symmetric background that acts as a simple dilution.  The different reconstructions are described in the text.}
\label{tab:ljetsTeV}
\end{table}

We can apply our simple global $\chi^2$ reconstruction to $l$+jets events with at least four jets, at least one of which must be $b$-tagged.  We use the basic analysis cuts of~\cite{Aaltonen:2012it}.  Normalizing to a partonic cross section of 7.4~pb, we predict about 2600 $t\bar t$ events, whereas CDF predicts 2186$\pm$314.  We scale our statistics to match CDF's central estimate.  In Tab.~\ref{tab:ljetsTeV}, we list the reconstruction-level \Aperpl\ induced by our set of axigluon models (under the ``$\ge 4j$ global" column).  The largest \Aperpl, about $4.7\%$, is from the 420~GeV ``light'' axigluon benchmark model, with 20\% width and couplings set to reproduce the central Tevatron \AFB.  Given the statistical uncertainties only, and neglecting backgrounds, this would constitute a $2.2\sigma$ effect.

However, since we are actually only interested in measuring the properties of the leptonic top, it is possible to relax the reconstruction requirements.  We have therefore further explored measurements in an event sample where we demand $\ge 3j$ instead of $\ge 4j$, increasing the number of events from 2200 to 3550.  Backgrounds, which were in 1:4 ratio with $t\bar t$ in the $\ge 4j$ sample, grow to about 1:1.5.  It is not clear to what extent the growth of backgrounds poses an obstacle to the measurement, since the intrinsic asymmetry of backgrounds such as $W+3/4j$ would need to be estimated with a high-multiplicity loop-level event generator.  We stress that high-multiplicity tree-level event generators, such as {\tt MadGraph}~\cite{Alwall:2011uj} or {\tt ALPGEN}~\cite{Mangano:2002ea}, are incapable of modeling the induced reconstruction-level \Aperpl, since it violates naive time-reversal symmetry (i.e., mirror reflection in the reconstructed production plane).  If the known results for $t\bar t$ can serve as a guide, then the background asymmetry might indeed be very small.  But the best that we can manage at this stage is to treat the backgrounds as a simple 60\% dilution factor of the measurable asymmetry, and a 66\% increase in sample statistics.  The net effect of this background is then a 20\% reduction in statistical significance of any asymmetry in $t\bar t$.

A very simple way to reconstruct the leptonic top would be to ignore the hadronic top entirely, looking for the $l\nu j$ grouping that comes closest to $m_t$.  But we have found that we can obtain approximately 20\% larger reconstruction-level asymmetries by instead keeping some amount of global event information.  Our leptonic top $\chi^2$ is defined as follows.  In events with $\ge 4j$ we perform the usual partitioning, and in events with $3j$ we only use two jets for the hadronic top.  If the hadronic-side mass is smaller than $m_t$, we simply define $\chi^2 = (m(l\nu j) - m_t)^2$.  If it is larger than $m_t$, we add in $(m(jj/jjj) - m_t)^2$.  In this way, {\it under}-reconstructions of the hadronic top are completely allowed, but partitionings that produce too-massive hadronic top candidates are penalized.  Tab.~\ref{tab:ljetsTeV} contains the reconstruction-level \Aperpl's obtainable with this approach (under the ``$\ge 3j$ $t_l$" column).   We have found that they are practically identical to the values obtained in the $\ge 4j$ subsample with the global $\chi^2$ reconstruction.\footnote{The breakdown into $3j$ and $\ge 4j$ subsamples using the $t_l$-based reconstruction is somewhat nontrivial.  For the light axigluon model, the asymmetries in the two samples are nearly identical, and in turn the same as the fully global $\ge 4j$ reconstruction.  For the heavy axigluon and contact-interaction models, the $t_l$ reconstruction's $\ge 4j$ asymmetry is about 30\% larger than the $3j$ asymmetry, and the fully global $\ge 4j$ reconstruction sits in between them.}  Accounting for the background, and assuming that it has zero intrinsic asymmetry, the central-coupling 420~GeV axigluon model would induce a $2.0\sigma$ effect with our previous global reconstruction (including the small backgrounds), and $2.2\sigma$ with the present leptonically-biased reconstruction.

It is possible to make further incremental improvements by demanding consistency between different reconstruction strategies.  As an entirely independent reconstruction appropriate to $\ge 3j$ events, we have also considered the best reconstruction of the {\it hadronic} top.  For this, we look for the grouping $j+jj$ that minimizes $(m(j+jj) - m_t)^2 + 4(m(jj) - m_W)^2$, completely ignoring the lepton and \met\ vector.  In $3j$ events there is only one possible grouping.  We assume that the $\vec{p}_T$ of the leptonic top is exactly back-to-back with the hadronic top candidate.  This reconstruction is not quite as faithful as the leptonic reconstruction.  (The central-coupling 420~GeV model's \Aperpl\ is $4.3\%$, compared to $4.7\%$.)  However, if we take only the subset of events where these two reconstructions yield the same sign($\cos\theta_y$), the measured asymmetries become significantly larger.  We also list these in Tab.~\ref{tab:ljetsTeV} (under the ``$\ge 3j$ $t_l$/$t_h$ agree" column).  The consistency demand shrinks the sample to 2440 events, and has little effect on $S/B$, but the increases in the asymmetries lead to a net gain in significance.\footnote{The proportion of $3j$ relative to $\ge 4j$ stays largely fixed.  For the light axigluons, \Aperpl\ becomes about 10\% larger for the $3j$ subsample relative to $\ge 4j$, whereas the reverse is true for the heavy axigluons and contact-interaction.}  For example, we achieve $2.6\sigma$ with the central-coupling 420~GeV axigluon.  The lower-coupling 420~GeV axigluon and central-coupling 800~GeV axigluon both give about $1.5$--$1.6\sigma$.  Therefore, using this reconstruction, a range of light axigluon models might be visible in $l$+jets at the Tevatron at the level of $2$--$3\sigma$.  As expected, heavy axigluons are more difficult.

We have also studied measurements using the dileptonic sample.  While the branching fraction is much lower than $l$+jets, and the kinematic reconstruction is complicated by the presence of two neutrinos, these disadvantages are partially offset by the fact that each event offers us two measurements of the individual top polarizations.  Still, a dileptonic measurement remains highly challenging, and it is not clear that it can be made to compete with $l$+jets.  We present some of our own observations in attempting to craft such a measurement, with the hope that these can serve to inform a more sophisticated analysis in the future.

The crucial point we make is that {\it the biggest obstacle to the analysis is not the missing neutrinos, but the correct pairing of b-jets and leptons}.  In fact, we could ignore the neutrinos entirely and simply build the two tops with the $b$-jet candidates and leptons.  Even accounting for the imperfect $b$-tagging and the fact that a spurious second ``$b$-jet'' can sometimes be provided by initial-state or final-state radiation, unambiguous charge-matching of the available tagged $b$'s and leptons would allow \Aperpl\ to be measured with greater than 2$\sigma$ significance for the central-coupling 420~GeV model.  By contrast, the maximum sensitivity that we were able to obtain with realistic pairing and candidate neutrinos built using information from \met\ and other event kinematics is only 0.8$\sigma$.
These results utilize the same dileptonic top reconstruction that we proposed in~\cite{Baumgart:2012ay}.  We demand one $b$-tag as part of the event selection, and the candidate jets for pairing are the tagged jet and the hardest remaining jet (or the other $b$-tagged jet in double-tagged events).  The full event selection is described in Appendix~\ref{sec:details}.  Backgrounds are small ($S/B > 10$), and we neglect them for our analysis.

After selecting two jets, there is a two-fold ambiguity in how to pair them with the two leptons.  We construct the tops' $m_{T2}$~\cite{Lester:1999tx,Barr:2003rg} for both possibilities and obtain individual neutrino transverse momenta from the numerical minimum.  Their longitudinal momenta are set to match the rapidities of their associated $b$+lepton four-vectors.  Pairing is done by first checking if $m_{bl}$ or $m_{T2}$ exceed kinematic constraints for either combination and discarding the offending solution.  If neither or both fail, then we take the combination that minimizes $(m(bl^+ \nu) - m_t)^2 + (m(\bar b l^- \bar \nu) - m_t)^2$.  For events in our Tevatron sample where both truth-level $b$-jets were reconstructed, this procedure has a 66\% chance to correctly pair.  For the central-coupling 420~GeV axigluon, the resulting \Aperpl\ is 2.7\% at 0.8$\sigma$, treating the $l^+$ and $l^-$ asymmetries as combined independent measurements.  Given the much better results obtainable in principle with a more idealized $b$-lepton pairing, we emphasize that a more sophisticated procedure aimed at improving this pairing would be worth pursuing.

\subsection{LHC}
\label{sec:LHC}

For $l$+jets at the LHC, we consider the 2012 data set of 20~fb$^{-1}$ at 8~TeV.  We use only $\ge 4j$ events.  We do not explore the possibility of $3j$ reconstructions, since we require a fully global picture of the event in order to measure $\Ytt$.  While we will see that we nonetheless obtain good statistical reach, we emphasize that an expanded analysis that includes $3j$ events would likely still be worth pursuing, especially since jets can overlap for the higher-$p_T$ regions of production phase space.  Approaches using jet substructure~\cite{Abdesselam:2010pt,Altheimer:2012mn} or non-isolated leptons might also be beneficial.

Our global event reconstruction is identical to the one that we used above for the Tevatron.  Besides the need to guess the correct beam direction, the major novelties with respect to the Tevatron are the much larger statistics and an appreciable fraction of events at partonic CM energies far above $t\bar t$ threshold.  These give us much more flexibility for zooming-in on regions of phase space where \Pperp\ and its associated \Aperpl\ are the largest.

\begin{table}[tp]
 \centering
\begin{tabular}{l||r|r|r||r|r|r}
LHC8, 20 fb$^{-1}$ & \multicolumn{3}{c||}{Global reconstruction} & \multicolumn{3}{c}{Perfect reconstruction} \\
$l$+jets           & \;Inclusive\;        & \;SR(light)\;         & \; SR(heavy)\;         & \;Inclusive\; & \;SR(light)\; & \;SR(heavy)\; \\ \hline \hline
\# $t\bar t$ events&  550k                 & 100k                  & 40k                    &  550k                  &  100k                 &  40k      \\
$S/B$              &  8                    & 9                     & 7                      &                        &                       &           \\ \hline\hline
420 GeV, central   &  0.5\% ($3.5\sigma$)  &  1.3\% ($3.9\sigma$)  & ``0''                  &  0.7\%                 &  2.0\%                & ``0''     \\ 
420 GeV, lower     &  0.3\% ($2.1\sigma$)  &  1.0\% ($3.0\sigma$)  & ``0''                  &  0.4\%                 &  1.3\%                & ``0''     \\ \hline
800 GeV, central   & -0.3\% ($2.1\sigma$)  & -0.6\% ($1.8\sigma$)  & -3.2\% ($6.1\sigma$)   & -0.5\%                 & -1.1\%                & -3.5\%    \\
800 GeV, lower     & -0.1\% ($0.7\sigma$)  & ``0''                 & -2.3\% ($4.4\sigma$)   & -0.3\%                 & -0.6\%                & -2.4\%    \\ \hline
contact, central   & ``0''                 & ``0''                 & -1.3\% ($2.5\sigma$)   & ``0''                  & ``0''                 & -1.3\%    \\
contact, lower     & ``0''                 & ``0''                 & -1.0\% ($1.9\sigma$)   & ``0''                  & ``0''                 & -1.0\%    \\
\end{tabular}
\caption{\it LHC (8~TeV, 20~fb$^{-1}$) reconstruction-level \Aperpl\ in $t\bar t$ events for our baseline models using our global reconstruction strategy, and with a perfect leptonic top reconstruction for reference.  We also indicate the statistical significance, assuming a symmetric background that acts as a simple dilution.  Entries labeled ``0'' indicate that asymmetries are consistent with zero at the level of our MC statistics (0.08\% for inclusive, 0.18\% for SR(light), and 0.29\% for SR(heavy)).}
\label{tab:ljetsLHC}
\end{table}

If we simply take an inclusive event sample, the induced asymmetries for all of our example models is well below 1\%.  The largest is for the central-coupling 420~GeV model, which reconstructs to 0.5\% (0.7\% if the leptonic top was perfectly reconstructed).  While the sample size is roughly 550k events, and the corresponding statistical significance is therefore better than $3\sigma$, we can easily enhance both the size of the asymmetry and its significance with additional kinematic cuts.  For the 420~GeV models, we apply a cut $|\Ytt| > 1.0$ (``SR(light)'').  For the 800~GeV models and contact interactions, we apply the following three cuts:  $\Mtt > 600$~GeV, $|\Ytt| > 0.5$, and $|\cos\Theta| < 0.7$ (``SR(heavy)'').  These cuts are only meant to be illustrative, as a realistic measurement, such as that for $A_C$, would likely be performed over several kinematic bins simultaneously.  We list our results in Tab.~\ref{tab:ljetsLHC}.  We see that all asymmetries can be enhanced to $>1\%$ magnitude with the different sets of cuts.  The central-coupling 420~GeV significance goes up to $4\sigma$.  The most difficult model, the low-coupling contact interaction, is at $2\sigma$.  Backgrounds, dominantly $W$+jets but with a relatively sizable $tW$ single-top contribution, are typically an order-of-magnitude smaller than the $t\bar t$ signal.  Their dilution of the statistical significances are about 5\%, and we incorporate this factor into our estimates.  Modulo possible systematic errors, which must be controlled at the sub-percent level, the effects of these models on \Aperpl\ in the $l$+jets channel would be readily measurable at the LHC.

In the dileptonic channel, we face the same challenges as at the Tevatron, as well as the complication of determining $\Ytt$ with two missing particles.  We apply the $m_{T2}$-based reconstruction strategy detailed in section~\ref{sec:Tevatron}.  The success rate for pairing $b$-jets with leptons, in events where both $b$'s are reconstructed is 66\%.  We find that this analysis gives us greatest sensitivity to the 800 GeV axigluon models.  In addition to the procedure and cuts described above, we also impose a variant of the ``SR(heavy)'' cut used for $l$+jets, but with $\Mtt > 600$~GeV, $|\Ytt| > 0.75$, and $|\cos\Theta| < 0.9$.  Combining measurements of \Aperpl\ for $l^+$ and $l^-$ gives \Aperpl\ = 2.2\% at 1.3$\sigma$ for the central-coupling 800~GeV model.  As at the Tevatron, it is possible that a more sophisticated pairing procedure, possibly involving a multivariate $\chi^2$ on the global final-state information, could improve the quality of the dileptonic measurement.

We have also explored measurements of the spin correlations of these models in the $l$+jets channel, following the suggestions of~\cite{Baumgart:2011wk,Baumgart:2012ay}.\footnote{The dileptonic channel would seem more suited to such a measurement, given the maximal analyzing power of leptons.  Our own studies suggest that this channel is less powerful than $l$+jets in probing our benchmark models, due to a combination of the lower statistics and the weaker ability to craft targeted kinematic cuts.}  To construct the correlation-sensitive observable, we take as our ``spin analyzers'' the lepton from the semileptonic top decay and the softer of the two non-$b$ jets as viewed in the hadronic top's rest frame.  The total spin analyzing power is approximately $0.5$.  (The $b$-jet itself can also serve as the spin analyzer, with somewhat smaller sensitivity.)  We measure the azimuthal angle of each particle about the $t\bar t$ production axis in the partonic CM frame, counterclockwise about the $t$ direction.  The transverse $y$-axis used for our polarization measurement defines $\phi = \pi/2$.  The {\it sum} of the two azimuthal angles is our correlation-sensitive observable, and exhibits a cosine-wave modulation that can be modified by the resonances (see Fig.~\ref{fig:phiSum}).  We obtain the most significant deviations for the central-coupling 800~GeV model.  Placing cuts of $\Mtt > 700$~GeV and $|\cos\Theta| < 0.7$, the reconstructed SM modulation amplitude is $-8.0\%$, and the resonance reduces this to $-4.8\%$.  This is approximately a 4$\sigma$ shift.  (The expected number of events passing the modified cuts is again close to 40k.)  The lower-coupling 800~GeV model and central-coupling contact-interaction lead to smaller shifts at approximately 2$\sigma$ significance.  The 420~GeV resonances would be much more difficult to see in this way, as their contributions to the differential rate at the peak are only a few percent, and the deviations in the spin correlations are less dramatic.  In general, we estimate that the transverse polarization effect is easier to see than the spin correlations for these broad axigluon models, though such a comparison would benefit from a more complete detector simulation and treatment of experimental and theoretical errors.

\section{Conclusions and Outlook}
\label{sec:conclusions}

The transverse polarization of individual tops at the Tevatron and LHC offers us a unique window into loop-level new physics.  In this paper, we have explored how the mechanism responsible for the anomalous Tevatron \AFB\ might also induce an anomalous transverse top polarization.  One of the few viable surviving explanations for \AFB\ is an axigluon-like resonance, which can remain hidden in the $t\bar t$ mass spectrum if it is very broad.  A large natural width indicates large loop-level corrections to the axigluon propagator, and automatically contributes a transverse polarization far in excess of that expected from QCD.

While the polarization effect is highly dependent on the axigluon couplings, mass, and width, we have studied a set of simple benchmark models that lead to observable effects.  For an axigluon not far above top pair threshold, the transverse polarization might be observable with almost 3$\sigma$ statistical sensitivity at the Tevatron.  Heavier axigluons, even in the contact-interaction limit, might be seen with anywhere between 2$\sigma$ and 6$\sigma$ statistical significance at the LHC.  If the Tevatron anomaly persists, transverse polarization can serve to verify that new physics is responsible.  Conversely, limits on the transverse polarization would help rule out a large class of axigluon explanations.

We hope that our paper has also highlighted the need for more rigorous predictions of the transverse polarization and its associated leptonic asymmetry in Standard Model $t\bar t$ and its backgrounds.  The ubiquitous tree-level event generators used to model $W$+jets do not capture this effect, and therefore require input from high-multiplicity NLO computations.  It will also be important to understand the electroweak corrections in $t\bar t$ production, as the QCD-only contributions are very small.

From a more general perspective, it is interesting to consider what else might be probed with this ``loop-level polarimeter.''  An obvious generalization is to use transverse polarization to search for broad scalar or spin-2 resonances in the $gg \to t\bar t$ spectrum.  Transverse polarization is also sensitive to the imaginary part of the top's anomalous chromomagnetic form factor, and many other non-resonant loop effects.  Understanding the LHC sensitivity to these scenarios would be a fruitful avenue for future work.


\acknowledgments{  We thank Martin Schmaltz and Gustavo Marques Tavares for useful discussions.
  MB was supported by DE-FG-03-91ER40682. 
  BT was supported by DoE grant No.\ DE-FG-02-91ER40676 and by NSF-PHY-0969510 (LHC Theory Initiative).}


\appendix

\section{Simulation Details}
\label{sec:details}

We generate the $t\bar t$ signal and its backgrounds at leading-order using \MadGraph5\ {\tt v1.4.7}~\cite{Alwall:2011uj} interfaced with \PYTHIA~\cite{Sjostrand:2006za}, and normalized to NLO.  To investigate the effects of axigluon resonances in fully-reconstructed samples, we apply event-by-event reweightings based on the 6-body final-state kinematics in $q\bar q \to t\bar t$ events in the parton-level event record.  We generate independent background samples for our studies in $l$+jets and dilepton decay modes, at both the Tevatron and the LHC.  All backgrounds are matched using MLM with five flavors, $R = 0.4$, and $p_T = 20$~GeV (25~GeV) for the Tevatron (LHC).  The $l$+jets Tevatron backgrounds include $W$+jets matched up to four jets and single-top matched up two jets ($t$- and $s$-channels).  The $l$+jets LHC backgrounds include $W$+jets matched up to four jets, single-top matched up to three jets ($t$- and $s$-channels), and $tW$ single-top matched up to one jet.  The dilepton Tevatron backgrounds before $b$-tagging are outlined for example in~\cite{CDF:2011afb}, with $W$+jets with fake leptons constituting the single largest contribution.  To obtain a rough estimate of the effect of $b$-tagging on the backgrounds, we simulate $W^+W^-$ matched up to two jets and normalized to match the rate reported by CDF.  We find that $S/B > 10$ would easily be achievable if the other backgrounds have tagging efficiencies even within the same order of magnitude.  For the LHC, we have checked  $l^+l^-$ (including $\tau^+\tau^-$) matched up to two jets, $W^+W^-$ matched up to two jets, and $tW$ matched up to one jet.  Again, we estimate $S/B > 10$ if a $b$-tag is applied.

After showering and hadronization in \PYTHIA, we process the particle output of the physics simulations into reconstructed leptons, jets, and \met.  We demand that leptons be isolated from surrounding activity within an $\eta$-$\phi$ cone of radius $R = 0.4$, such that the scalar-summed $p_T$ of the cone particles cannot exceed 10\% of the lepton's $p_T$.  Tevatron leptons should have $p_T > 20$~GeV and $|\eta| < 1.0$, and LHC leptons should have $p_T > 30$~GeV and $|\eta| < 2.5$.  (Leptons that fail these criteria are treated as ``hadrons'' and clustered into the jets.)   The remaining particles in the event we cluster into $R = 0.4$ jets in \FastJet\ {\tt v2.4.2}~\cite{Cacciari:2005hq}, using the JETCLU algorithm at the Tevatron~\cite{JETCLU} and the anti-$k_T$ algorithm~\cite{Cacciari:2008gp} at the LHC.  We keep Tevatron jets with $p_T > 20$~GeV and $|\eta| < 2.0$, and LHC jets with $p_T > 30$~GeV and $|\eta| < 2.5$.  We determine whether a jet carries flavor by looking back through the \PYTHIA\ event record for the hardest bottom- or charm-hadron within the jet radius, not counting charm generated in bottom decay.  Each jet then has some probability of being $b$-tagged.  Tevatron tag rates for ($b$,$c$,light) jets are assumed to be (40\%,6\%,1\%), and LHC tag rates (70\%,10\%,2\%).  For both $l$+jets and dileptonic events, and at both colliders, we demand at least one $b$-tagged jet.  Additionally, for dileptonic events at the Tevatron we require \met\ $>$ 25 GeV and $H_T  >$ 200 GeV, and at the LHC we require \met\ $>$ 30 GeV.

To roughly model detector energy resolution, we smear the energies of reconstructed leptons and jets before the application of acceptance cuts.  At the Tevatron, we use $\sigma(E)/E = (0.135)\sqrt{{\rm GeV}/p_T} \oplus 0.02$ for electrons, $(0.001)p_T/{\rm GeV}$ for muons, and $\sigma(E_T) = (0.1)E_T + (1.0\; {\rm GeV})$ for jets~\cite{Acosta:2004hw}.  At the LHC, we use $\sigma(E)/E = 0.02$ for electrons, $(0.1)\sqrt{E/{\rm TeV}}$ for muons, and $(0.8)\sqrt{{\rm GeV}/E} \oplus 0.04$ for jets.  At the LHC, we further smear the directions of the jets following~\cite{CMSpf}, by $0.025$ separately in $\eta$ and $\phi$.\footnote{This jet energy/direction smearing adds to effects already introduced by parton showering and jet reconstruction.  The direction smearing roughly models the spatial effects of the LHC detectors.  We do not smear jet directions for the Tevatron, since the JETCLU algorithm in \FastJet\ already applies a discrete calorimeter model.}  We define \vecmet\ to balance the vector-summed transverse momentum of all reconstructed leptons and jets.


\bibliography{lit}
\bibliographystyle{apsper}

\end{document}